\documentstyle[12pt,psfig]{article}
\def\aprle{\buildrel < \over {_{\sim}}} 
\def\aprge{\buildrel > \over {_{\sim}}} 

\begin{document}     
\title{Comparison of $\nu_\mu \leftrightarrow  \nu_\tau$ 
and  $\nu_\mu \leftrightarrow \nu_s$ oscillations  as
solutions of the atmospheric neutrino problem}
\author{Paolo Lipari and Maurizio Lusignoli \\
Dipartimento di Fisica, Universit\`a di Roma ``la Sapienza",\\
and I.N.F.N., Sezione di Roma, P. A. Moro 2,\\ I-00185 Roma, Italy}

\date{~~~~}

\maketitle
\begin{abstract}
The simplest explanation for the atmospheric neutrino anomaly
is the transition of muon neutrinos and antineutrinos into
states of different flavor.  Data  from the reactor experiment
Chooz, and the measurement  of the energy and  angle  distribution
of electron--like events in Super--Kamiokande put strong  constraints
on $\nu_\mu$--$\nu_e$ mixing, but 
the transitions $\nu_\mu \to \nu_\tau$ and $\nu_\mu \to \nu_s$ 
are  viable  possibilites.  If the muon neutrinos
are mixed with a singlet sterile state,  the  
energy and  zenith angle dependence of the oscillation probability
is  modified by matter effects, that
are important for  $E_\nu/|\Delta m^2| \aprge 10^3$~GeV/eV$^2$.
We confront the theoretical predictions with existing data on contained
events and neutrino--induced upward going muons.
\end{abstract} 


\vspace{\baselineskip}

\section{Introduction}
Recent results 
on atmospheric neutrinos \cite {SK-sub,SK}
show a deficit in the number of $\nu_\mu$--induced events 
that can be interpreted as a consequence of neutrino oscillations.
Additional signals of $\nu$--oscillations  come  from solar neutrinos
\cite{solar} and  from the LSND experiment \cite {LSND}.
These  results, if all interpreted 
in terms of  oscillations, require three independent
values of the parameter $\Delta m^2$, seem  to imply the
existence of a least four neutrino  mass  eigenstates \cite{three-neutrinos}
and  therefore, given the limits on the number  of flavors of standard
neutrinos \cite{LEP}, the existence  of at least one sterile state.

The energy and  zenith angle  distributions of the 
electron--like  events  detected  by Super--Kamiokande 
\cite{SK} are in agreement  with the Montecarlo
prediction without oscillations;
at the same time, the results of the Chooz experiment
\cite{Chooz}  put  stringent  limits on  
$\nu_\mu$--$\nu_e$ mixing 
in the region of $\Delta m^2$ of interest,
therefore  the hypothesis  of $\nu_\mu \leftrightarrow \nu_e$ 
oscillations  as explanation of the atmospheric neutrino data
is disfavored  (see 
\cite{Concha98} and \cite {Foot-Volkas-numu-nue} for 
a more quantitative discussion).
The atmospheric  neutrino data 
can be explained  by the existence of   
$\nu_\mu \leftrightarrow \nu_\tau$ or $\nu_\mu \leftrightarrow 
\nu_s$ oscillations,
where $\nu_s$ is a sterile neutrino.
The possibility that  the atmospheric   neutrino anomaly is
due to $\nu_\mu$--$\nu_s$ mixing has been   previously discussed 
in  \cite {ALL93},  and  more recenty 
in \cite{ster_other,Vissani-Smirnov,Liu-Smirnov,Foot_etal}.  

Bounds on the existence of sterile states with large mixing 
to the standard neutrinos have been  obtained
from  cosmological  considerations \cite{cosmological-bounds}.
For large enough  values of the mixing angle and $\Delta m^2$,
the oscillations of standard  into sterile neutrinos 
can  bring the sterile states in (or close to) thermal 
equilibrium with matter before the nucleosynthesis epoch.
The  increase in  the energy density  
results in  the overproduction of primordial helium, 
spoiling  the success of Big Bang Nucleosynthesis.
Schramm and Turner  \cite{Schramm-Turner}, 
have  estimated   the upper limit 
(at 95 \% C.L.) on the number of light degrees of freedom
at  nucleosynthesis  as  $N_\nu \le 3.6$.
This  bound, according to  the analyses  in \cite{cosmological-bounds},
is incompatible with the  explanation of  the atmospheric neutrino problem
as due to the mixing betwen  muon and  sterile neutrinos, because the 
allowed  region in the oscillation 
parameter  space  would  result in $N_\nu \simeq 4$.
However recent  work  \cite{Foot-Volkas}  has  shown 
that this  cosmological bound  can be 
evaded,  considering   the suppression of  oscillations
due to the  possible  presence of a lepton asymmetry in the early universe
that can be generated  by the oscillations themselves.

It is interesting to investigate  if the data on atmospheric  neutrinos
already existing  or to  be collected in the near future  can 
discriminate  between these two different hypotheses for oscillations.
The energies   of atmospheric  neutrinos  are such that 
only  a very  small  fraction is  above the threshold
for  $\tau$--lepton charged current (CC) production, on the other hand 
the neutral current (NC) cross sections  of $\nu_\tau$ and $\nu_\mu$ are 
identical.
Sterile  neutrinos instead, by definition, do not interact even  by neutral 
currents. In principle  the  presence  of a flux  of tau  neutrinos
can  therefore  be identified  as an excess of NC over CC events.
The  measurement is possible in large mass  detectors like 
Super--Kamiokande (SK). The most  promising       method,
as  recently  discussed
by Smirnov and Vissani \cite{Vissani-Smirnov},
is  the study of the reaction $\nu_x N \to \nu_x N \pi^\circ$
where  a single $\pi^\circ$ is  produced, and  detected   after  its  decay
$\pi^\circ \to \gamma\gamma$.
The absolute value  of the  cross section for this  process 
is not known  with great  accuracy,    because
it has  only  been measured with  large errors \cite{NC-exp} 
and it is difficult to  compute  theoretically \cite {NC-theo}.
Part of this   uncertainty   should cancel \cite{Vissani-Smirnov}
when comparing  with the  rate of single  pion  events  
produced in charged--current interactions.  Experimentally 
one will need to  know with sufficient 
accuracy the   acceptance and efficiency  for the  detection of events 
with at least two `tracks' in the final state.

Another  handle to  distinguish  the
$\nu_\mu$--$\nu_\tau$ and $\nu_\mu$--$\nu_s$  mixing 
is given by the fact that the energy and zenith angle  dependence
of the oscillation probabilities are  different  in the  two  cases
because of the presence of   matter effects \cite{MSW}.
In this work we discuss  in detail this  possibility, examining the 
effects that could be observed both in the events induced by neutrino
interactions in the detector and in the upward going muons, generated by 
neutrino interactions in the rock beneath the detector. As we will show, 
the most promising tool to 
disentangle matter effects should be provided by the study of upward going
muons, notwithstanding the present confusing experimental situation.

\section {Matter  effects and oscillation probabilities}

\subsection {Oscillation parameters in matter}
The $\nu_\alpha \to \nu_\beta$ oscillation  probability in vacuum,
in the simple case of two flavor  mixing, is
given  by the standard  formula:
\begin{equation}
P_{\nu_\alpha \to \nu_\beta} = \;
\sin^2 2\,\theta ~\sin^2
\left  ( { \Delta m^2 \, L \over 4 E_\nu} \right )
\equiv 
\sin^2 2\,\theta ~\sin^2 \left (\pi\;{L \over \ell} \right ) 
\label{prob-vacuum}
\end{equation}
where  $\theta$, 
defined in the interval $[0,\pi/4]$,  
 is the mixing angle that  relates the   flavor  and mass eigenstates
($|\nu_\alpha\rangle = \cos\theta\, |\nu_1\rangle + \sin\theta\,|\nu_2\rangle$,
$|\nu_\beta \rangle = -\sin\theta\,|\nu_1\rangle + \cos\theta\,|\nu_2\rangle$)
and $\Delta m^2 = m_2^2 - m_1^2$, with  $m_1$ and $m_2$ the mass eigenvalues.

In the presence of  matter with  constant density
the  mixing angle $\theta$ and  the oscillation length $\ell$
must be replaced by $\theta_m$ 
and $\ell_m$, defined as follows:
\begin{equation}
\sin^2 2\,\theta_m ={\sin^2 2\,\theta \over
[\zeta  -\cos 2\,\theta]^2+ \sin^2 2\,\theta},
\label {mix-mat}
\end{equation}   
\begin{equation}
\ell_m = {\ell \over
\{ [\zeta -\cos 2\,\theta]^2+  \sin^2 2\,\theta \}^{1/2} }.
\label{losc-mat}
\end{equation}   
where 
\begin{equation}
\zeta = {2 \,E_\nu \,V_{\alpha\beta} \over \Delta m^2}
\label{eq_zeta}
\end{equation}
and  $V_{\alpha\beta} = V_\alpha - V_{\beta}$ is the difference
between  the effective potentials in matter of  
neutrinos  of flavor $\alpha$ and $\beta$  \cite{MSW}.
The effective potentials of $\nu_\mu$ ad $\nu_\tau$ in ordinary matter
are  identical,  and  matter effects    vanish  in the 
case of $\nu_\mu$--$\nu_\tau$ mixing.
In the case of $\nu_\mu$--$\nu_s$ mixing  the   difference in the effective
potential is
\begin{equation} 
V_{\mu s} = \mp\sqrt{2}\,G_F\, {N_n \over 2} 
= \mp 1.9 \cdot 10^{-4}
~{{\rm eV}^2  \over {\rm GeV} }
~~\left ( {\rho \over \rho_\circ } \right )
\left (1 - {Z \over A} \right ),
\label {eq:potential}
\end{equation}
where the minus (plus)  sign is  for neutrinos (antineutrinos),
$G_F$ is the Fermi  constant,
$N_n$  the neutron number density, $\rho$ the mass density
($\rho_\circ$ = $5~{\rm g~cm}^{-3}$ is a reference value)
and $(1-Z/A)$ the neutron fraction. 
For  comparison we note  that the effective  potential  
in the case of $\nu_e$--$\nu_\mu$ mixing
is $V_{e \mu} = \pm \sqrt{2}G_F\,N_e$  where $N_e$ is the electron  density;
in the earth $N_e \simeq N_n$ and  $V_{e \mu} \simeq -2\, V_{\mu s}$.

\begin{figure} [t]
\centerline{\psfig{figure=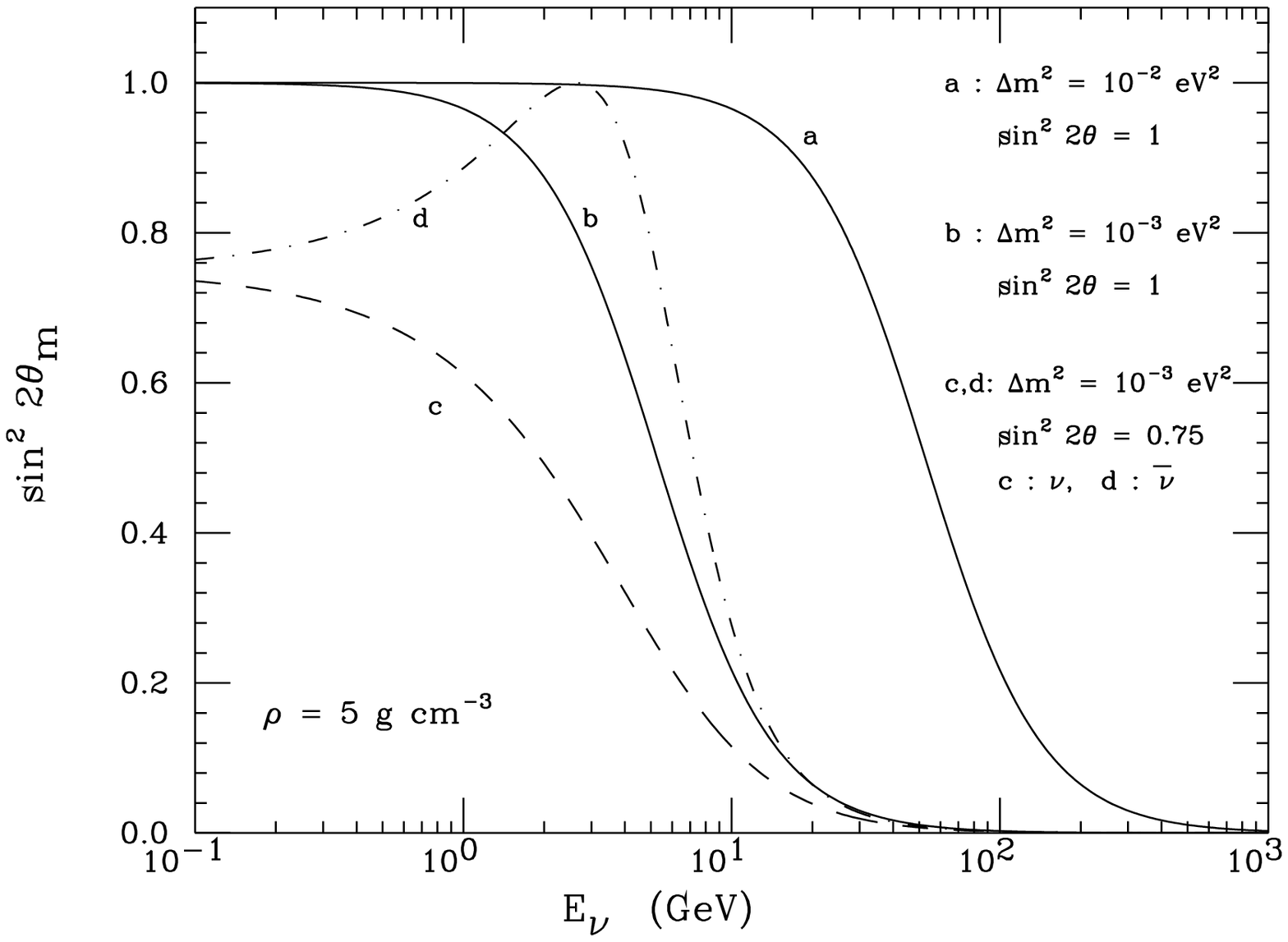,height=9cm}}
\vspace {0.4 cm}
\caption{Effective mixing parameter for $\nu_\mu$--$\nu_s$ oscillations
in matter  with  constant density $\rho = 5$~g~cm$^{-3}$ and  neutron
fraction 0.5.
\label{fig_osc1} }
\end{figure}

\begin{figure} [t]
\centerline{\psfig{figure=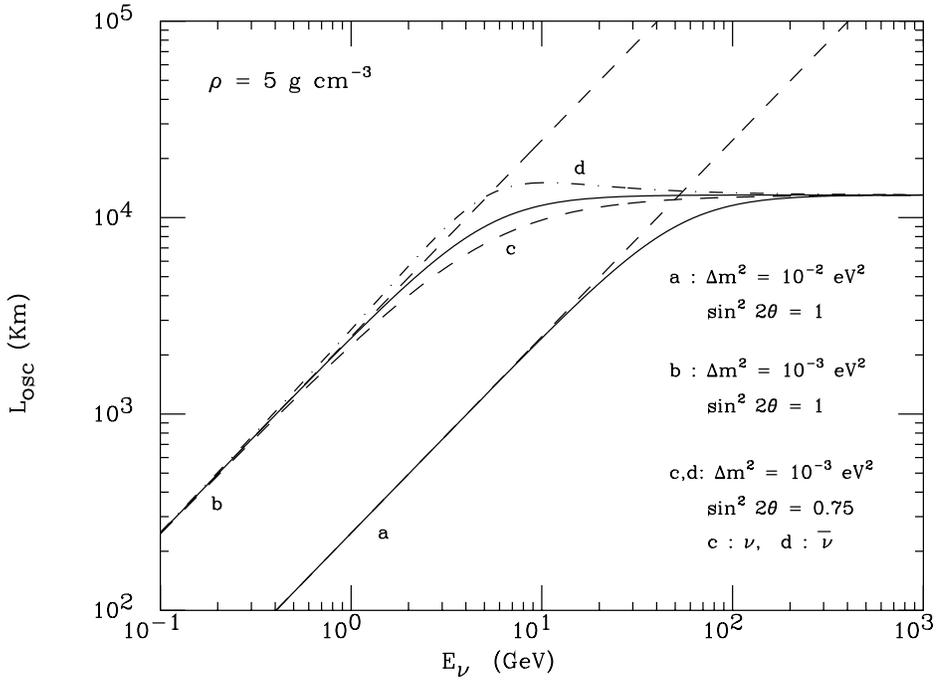,height=9cm}}
\vspace {0.4 cm}
\caption{Oscillation length in  matter.
\label{fig_osc2} }
\end{figure}

The meaning of equations (\ref{mix-mat}) and (\ref{losc-mat}) is illustrated 
in figs.~\ref{fig_osc1} and~\ref{fig_osc2}, respectively. In these 
figures we 
have plotted the oscillation parameters in matter for a  fixed value
of the density $\rho=\rho_\circ$ and for values of $\Delta m^2$ and 
$\sin^2 2\,\theta$ relevant for atmospheric neutrinos.
The matter effects are controlled by 
the ratio $\zeta$ defined in eq.~(\ref{eq_zeta}).
For $|\zeta| \ll 1$   (small neutrino energies)
the matter effects  are  negligible, and  the oscillations
proceed as  in vacuum.
For $|\zeta| \gg 1$   (large neutrino energies)
the matter effects  are dominant and 
the oscillations are strongly supressed: the
effective mixing   $\sin^2 2\,\theta_m$ decreases like 
$\zeta^{-2}$ and the oscillation length levels off to a value 
$\ell_m^\infty = 2\,\pi\,/\,|V_{\mu s}|
 \simeq 1.3 \,(\rho_\circ/\rho)\;10^4$~km, independent from 
$E_\nu$ and $\Delta m^2$, and remarkably
close to the earth's diameter.
In the region $|\zeta| \sim  1$,  that corresponds to 
a neutrino  energy
\begin{equation}
E_\nu   \sim { |\Delta m^2| \over \sqrt{2} G_F \, N_n } = 5.2~ {{\rm GeV}} ~~
 \left ( {|\Delta m^2| \over 10^{-3}~\rm {eV}^2} \right )
 ~\left ( { \rho_\circ \over \rho} \right )\;,
\label{E_typical}
\end{equation}
one has the most complex behavior.
In general, with the exception of the special case of maximal mixing
($\sin^2 2\,\theta$ = 1),
neutrinos  and  antineutrinos have different  oscillation parameters
in matter, because of the opposite  sign of the effective potential.
The effective  mixing in matter  becomes   maximal  
(MSW resonance effect \cite{MSW}) when the  condition
$\zeta = \cos 2 \theta$ is satisfied.  This   is  possible for 
antineutrinos (neutrinos) 
when $\Delta m^2$ is positive (negative) and is  illustrated  by  the curves  
labeled `c' and `d'.

Inspection of figs.~\ref{fig_osc1} and~\ref{fig_osc2}
 allows to understand qualitatively
when the matter effects can be important, 
and  the  hypotheses of $\nu_\mu \leftrightarrow \nu_\tau$ 
and $\nu_\mu \leftrightarrow \nu_s$  
oscillations experimentally distinguishable.
Atmospheric  neutrinos have  been observed
in essentially three energy regions: (a) the sub--GeV region:
events with charged lepton energy in the   range 
0.1--$0.2 \le E_\ell \le 1.33$~GeV,  produced  by 
neutrinos  with average energy $\langle E_\nu \rangle \simeq 0.7$--0.8~GeV;
 (b) the multi--GeV region:
events with charged lepton energy in the  interval 
$E_\ell \ge 1.33$~GeV,  produced by   neutrinos
with $\langle E_\nu \rangle \simeq 7$~GeV; 
(c) upward--going muons,  produced by  a broad  range of  neutrino  energies 
with a  median  energy of order 100~GeV.
The analyses of the sub--GeV  and multi--GeV  data in terms of 
neutrino oscillations suggest for $|\Delta m^2|$ a value 
in the range $10^{-3} \div 10^{-2}$~eV$^2$, therefore 
matter  effects  are essentially negligible in 
the  sub--GeV region, they can  be  relevant 
in the multi--GeV  region
if $|\Delta m^2|$ is close to the lower end of the above range, 
and are always significant for upward  going muons.

\subsection {Oscillation Probabilities}

In the earth
the composition remains always  close to the value $Z/A \simeq 0.5$,
but the density is  not  constant growing from 
a  value of  $2.7$~g~cm$^{-3}$ at the surface to $12.5$~g~cm$^{-3}$
close to the center. Therefore the effective oscillation
parameters are not constant, 
and the flavor conversion probabilities  must be 
obtained integrating the flavor evolution equation.

\begin{figure} [t]
\centerline{\psfig{figure=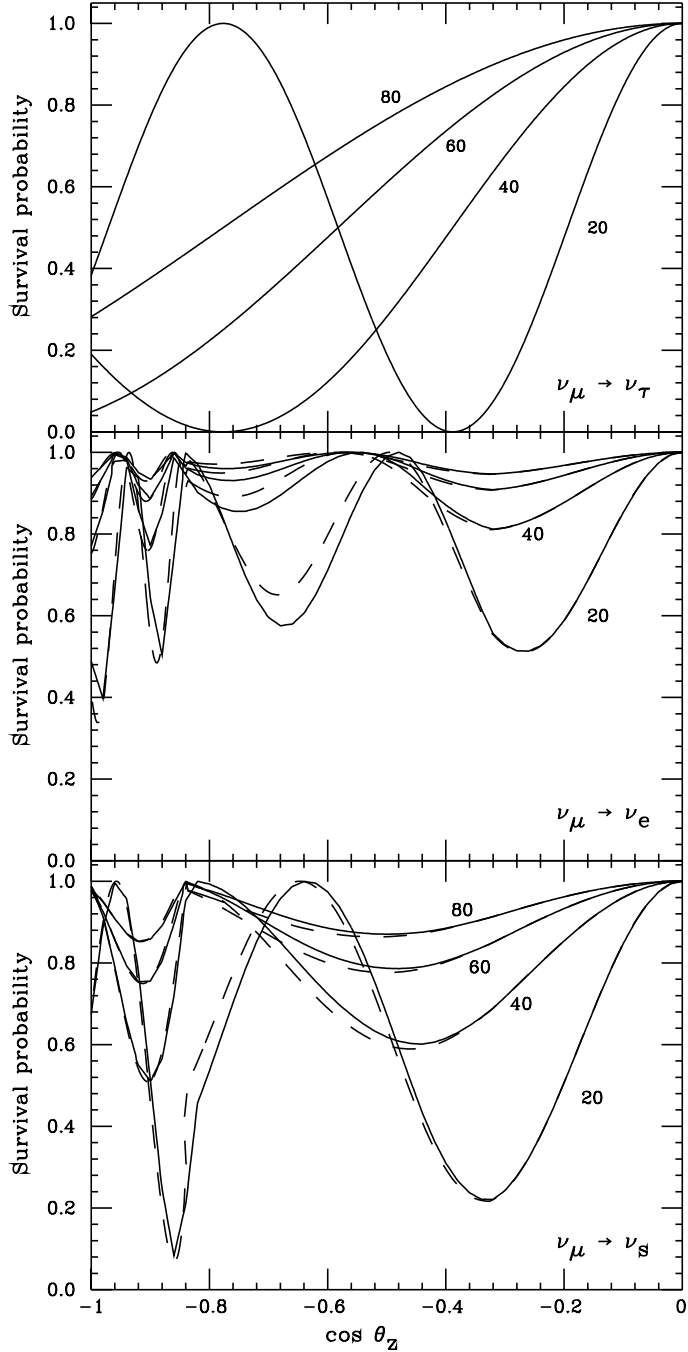,height=18cm}}
\vspace {0.4 cm}
\caption{Survival  probability 
$P(\nu_\mu \to \nu_\mu)$  as  a function of  the zenith angle
in the case of maximal mixing of $\nu_\mu$
with $\nu_\tau$ (upper panel), $\nu_e$ (middle panel)
and $\nu_s$ (lower panel). 
For $|\Delta m^2| = 5\cdot 10^{-3}$~eV$^2$ the curves correspond to
neutrino energies 20, 40, 60 and 80 GeV.
The dashed curves are calculated with the approximation of
constant average densities in the mantle and in the 
core of the earth \protect\cite{footnote-density}.
\label{fig_prob_a} }
\end{figure}

In  fig.~\ref{fig_prob_a} we plot 
the survival probability  of muon  neutrinos  
as  a function of the zenith angle, in  the case
of maximal mixing with  tau, electron and sterile  neutrinos.
We are assuming that the neutrinos  are created  
in a pure $\nu_\mu$ state at the surface 
of the earth, and observed when  they   emerge again from  the earth
after traveling a  distance $L = -2~R_\oplus\cos\theta_z$
($R_\oplus \simeq 6371$~km is the earth  radius).
In each panel  the curves correspond to  the survival probability 
of a neutrino with energy $E_\nu = 20$, 40, 60 and 80 GeV
if $\Delta m^2 = 5 \cdot 10^{-3}$~eV$^2$. The oscillation probability
is a function of $E_\nu/\Delta m^2$ and  the same  curves  
describe the probabilities for different values
of $\Delta m^2$ with a suitable rescaling of the values of the energy.

The upper panel shows the vacuum  probabilities (appropriate for 
$\nu_\mu$--$\nu_\tau$ mixing), that
are  sinusoidal  curves of constant  amplitude  
and oscillation length linearly growing with $E_\nu$.
In the middle and
lower panel the  survival 
probability for $\nu_\mu \leftrightarrow \nu_e$  and
$\nu_\mu \leftrightarrow \nu_s$ oscillations are plotted.
The most   remarkable feature to note is  the fact that 
for  large  $E_\nu$ the curves take  an  approximately
constant shape, with  minima 
at  fixed values of the zenith angle, 
but with an amplitude that decreases as $E_\nu^{-2}$.
In  the case of $\nu_\mu \leftrightarrow \nu_e$ oscillations, the
asymptotic  shape of the 
transition  probability has   four maxima  at
$-\cos\theta_z \simeq 0.32$, 0.76, 0.90 and 1.
In the case of oscillations into sterile  neutrinos
the large energy transition probability   
has  two maxima, a broad one  at $\cos \theta_z \simeq -0.45$ and a 
sharper and slightly higher one
at  $\cos \theta_z \simeq -0.90$~\cite{footnote-parametric}.  
We can  note that  in the lower panel
the   curve  labeled  20, that refers   to  a value
$E_\nu/\Delta m^2 = 4,000$~GeV/eV$^2$, is still of  an `intermediate'
shape, between the vacuum oscillation form and the asymptotic  shape
of the highest  energies; in the $\nu_\mu \to \nu_e$  case
where  the matter  effects  are stronger ($V_{e \mu} = -2 V_{\mu s}$)
the corresponding curve has
a shape closer to the asymptotic one.

To understand  qualitatively these  patterns  in the
oscillation probability  it is useful   to consider briefly
propagation in a medium with constant  density and composition.
In this ideal case the  oscillation  probability 
can be written as:
\begin{equation}
P_{\nu_\alpha \to \nu_\beta}  =
\sin^2 2 \theta ~{\epsilon^2 \over 
(1 - \epsilon \cos 2 \theta)^2 + \epsilon^2 \sin^2 2 \theta}
~\sin^2 \left [ {V_{\alpha \beta} L \over 2} \,
\sqrt {(1 - \epsilon \cos 2 \theta)^2 + \epsilon^2 \sin^2 2 \theta}
\right ]\;,
\label {p-matter}
\end{equation}
with  $\epsilon = \zeta^{-1} = \Delta m^2/(2 E_\nu V_{\alpha \beta})$.
In the  limit of  large neutrino  energy 
(that is for $\epsilon \to 0$) 
the transition probability  becomes:
\begin{equation}
\lim_{E_\nu \to \infty} P_{\nu_\alpha \to \nu_\beta}^{\rm matter}  = 
\sin^2 2 \theta ~{|\Delta m^2|^2  \over 4\,E_\nu^2 V_{\alpha \beta}^2 }
~\sin^2 \left [ {V_{\alpha \beta} L \over 2} \right ] \,,
\label{prob-asymp-matter}
\end{equation} 
that oscillates  with an asymptotic  oscillation length  
$\ell_{m}^{\infty} = 2\pi/|V_{\alpha \beta}|$ 
independent  from $\Delta m^2$ and
$E_\nu$, and  only  a  function of the density and  composition 
of the  matter, and with  an amplitude 
decreasing  with the energy  as $E_\nu^{-2}$.
These  are   qualitatively exactly the main features 
present  in the  middle  and  lower panels  of 
fig.~\ref{fig_prob_a}.

For  a fixed value of $L$, the large energy limit   of the oscillation
probability in vacuum is  readily obtained from eq.~(\ref{prob-vacuum}) as:
\begin{equation}
\lim_{E_\nu\to \infty} P_{\nu_\mu \to \nu_\tau}  =
\sin^2 2 \theta ~{|\Delta m^2|^2 L^2 \over 16 \,E_\nu^2 }\;.
\label{prob-asymp-vacuum}
\end{equation}
Comparing  equations (\ref{prob-asymp-matter})  and
(\ref{prob-asymp-vacuum}), we  note 
that they  differ because in the presence of matter the dependence
on the distance $L$ has a 
characteristic oscillating form instead of a quadratic increase.
In the case of $\nu_\mu$--$\nu_s$ mixing
the asymptotic oscillation  length
varies between  23,000~km  (at the surface)
and 5,300~km  (at the earth's center),  that corresponds to 
3.65 and 0.81 earth radii, respectively;
for $\nu_\mu$--$\nu_e$,  $\ell_{m}^{\infty}$ is a factor of  two shorter.
This  remarkable  coincidence between the 
oscillation  length  of high energy $\nu_\mu$'s mixed
with electron or sterile  neutrinos  when 
traversing the earth and the geometrical  radius of our planet,
generates  a characteristic  and  observable deformation of the zenith 
angle distribution of the high energy neutrinos coming from below the horizon.

The  considerations  that we have outlined   have  been  developed 
for constant density,  but the  qualitative  conclusion
that  for large $E_\nu$ the curve of the oscillation probability
as  a function  of zenith angle reaches  an  asymptotic  shape
remains  valid also in the  presence of
variable  density and  composition. In this  more general case
the shape is not  exactly sinusoidal;  for example one  can see
from fig.~\ref{fig_prob_a}  that the  maxima of the
oscillation probability   become  narrower for neutrino  trajectories
that  pass close to the center of the earth where the
density is  higher  and the oscillation length shorter. 
 We note  that  the  density of the earth 
has a sharp  discontinuity
at a radius $r_c = 3493$~km, the  boundary between the mantle and
the core, where the density jumps from 5.5  to 9.9~g~cm$^{-3}$ \cite{Stacey}.
A  simple  description of the density of the earth
as  composed  of two  layers of  constant density
with a  discontinuity at $r_c$, is  adequate 
to  describe most of the features of the oscillation 
probabilities, as  shown   by the dashed  curves \cite{footnote-density}
in the figures.

\begin{figure} [t]
\centerline{\psfig{figure=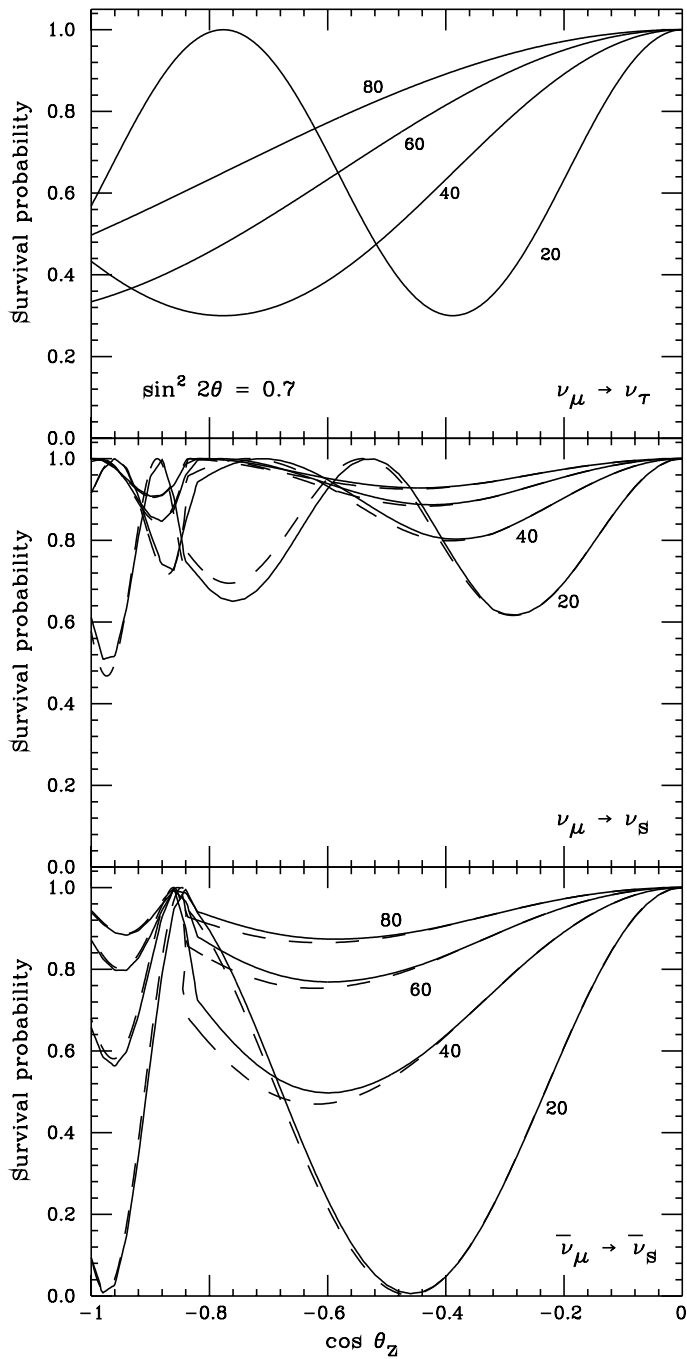,height=18cm}}
\vspace {0.4 cm}
\caption{As in figure \protect\ref{fig_prob_a}. 
The  mixing parameter is $\sin^2 2 \theta = 0.7$.
\label{fig_prob_b} }
\end{figure}

In fig.~\ref{fig_prob_b}
we  show survival probability curves for the case of
a vacuum mixing angle with $\sin^2 2 \theta = 0.7$.  The upper
panel  is for $\nu_\mu \leftrightarrow \nu_\tau$ oscillations,  
the middle (lower) panel
describe $\nu_\mu \leftrightarrow \nu_s$ 
($\overline{\nu}_\mu \leftrightarrow \overline{\nu}_s$) oscillations
for $\Delta m^2 >0$. 
An MSW resonance is present for the antineutrinos
when the condition $2\,E_\nu (-V_{\mu s})/\Delta m^2 =  \cos 2 \theta$ is
satisfied,  
while  the mixing in matter  of  the neutrinos  is  always suppressed
with respect to the vacuum case.
In the case $\Delta m^2 <0$  one should
exchange the  oscillation probabilities of neutrinos  and antineutrinos.
We can  also  note  that asymptotically the shape
of the   oscillation probability as a function of
zenith angle is the same  as in the maximal mixing  case,
however at all energies the oscillation length   
of neutrinos (antineutrinos)  is shorter (longer) 
than the maximal  mixing case.
The  detection of the angular  dependence of the
oscillation probability is therefore  more
difficult  if neutrino and antineutrino events are summed together.

\section {Neutrino interactions in the detector}
We  discuss first the information that can be obtained  from the events
where  the neutrinos interact  in the detector  via  charged  currents.
In  the  presence  of $\nu_\mu \leftrightarrow \nu_\tau$ 
or $\nu_\mu \leftrightarrow \nu_s$ 
oscillations, the rate of muon--like  events is  reduced,  and its  zenith 
angle  and  energy   dependence   is  deformed. 
The  absolute value  of the theoretical prediction \cite{nu-flux-comp}
has  a large  ($\sim 30$\%) uncertainty 
that    however  to a good  approximation affects  in
equal  way the   muon and  electron (anti)neutrino  fluxes, and  therefore
cancels  in the analysis  of the `double ratio' $R =
(N_\mu/N_e)_{DATA}/(N_\mu/N_e)_{MC}$.
The   Super--Kamiokande collaboration    \cite {SK} 
 in an  exposure of 25.5 kton~yrs, 
has  measured   
$R = 0.610^{+0.029}_{-0.028} \pm 0.049$  for  the sub--GeV sample
and
$R = 0.659^{+0.029}_{-0.058} \pm 0.053$  for  the multi--GeV, showing  
a significant   deviation   from   the expected value of 1.    Fitting  the
energy and zenith angle  dependence  of the  $e$--like  and $\mu$--like 
events  (the details  are  given in \cite{kamioka-semi,SK}),  the 
SK collaboration has  calculated  an allowed region at 90\% C.L. in
the ($|\Delta m^2|$, $\sin^2\,2\theta$) plane for  $\nu_\mu$--$\nu_\tau$ 
oscillations.
It is  in principle  straghtforward to  repeat  the  analysis
for the case of $\nu_\mu$--$\nu_s$  mixing.
Recently Foot, Volkas and Yasuda \cite{Foot-Volkas-numu-nue,Foot_etal}
have performed precisely this program,   using the  measurements
of $R$  and of the angular  distributions  of $\mu$--like and
$e$--like events to   recalculate the allowed  region  for the
$\nu_\mu$--$\nu_\tau$   hypothesis,  estimating  also 
the allowed  regions for 
$\nu_\mu$--$\nu_e$ and   $\nu_\mu$--$\nu_s$   mixing.
The result  obtained  in \cite{Foot_etal}   is that the
regions allowed for  $\nu_\mu$--$\nu_\tau$  and
$\nu_\mu$--$\nu_s$  oscillations  are  very similar, with the  second one  only
slightly  smaller, because the lowest
values of $|\Delta m^2|$ allowed   for  oscillations 
into $\tau$--neutrinos  are excluded in the sterile 
case~\cite{footnote-Foot}.
In this  work  we will not  attempt to  calculate  
again an allowed  region in  parameter  space for $\nu_\mu$--$\nu_s$ 
mixing.   We  find that the result of \cite{Foot_etal} is  qualitatively
correct. For a more detailed  analysis, one woud  need 
to have further informations on the detector  
acceptance and  efficiency,  and
a  knowledge  of the  bidimensional distribution in  energy and  direction of 
the events,  that has not been published.   

We would like  instead  to  argue 
that the analysis  of the zenith angle  and  energy 
dependence of the suppression of 
the $\mu$--like events  produced in the detector
can  distinguish  between 
the  hypotheses  of oscillations
into tau  or sterile neutrinos  
only  for $|\Delta m^2| \aprle 4 \cdot 10^{-3}$~eV$^2$.
This  follows  from the  facts  that:
(a)  the  matter   effects, depending  on the
quantity $E_\nu\,\rho/\Delta m^2$,  vanish  at  low energy;
(b) the  atmospheric  neutrino flux  falls steeply   with  energy,
and the rate  of interactions  of  high energy    neutrinos is
small even  for the large  mass of  Super--Kamiokande.

\begin{figure} [t]
\centerline{\psfig{figure=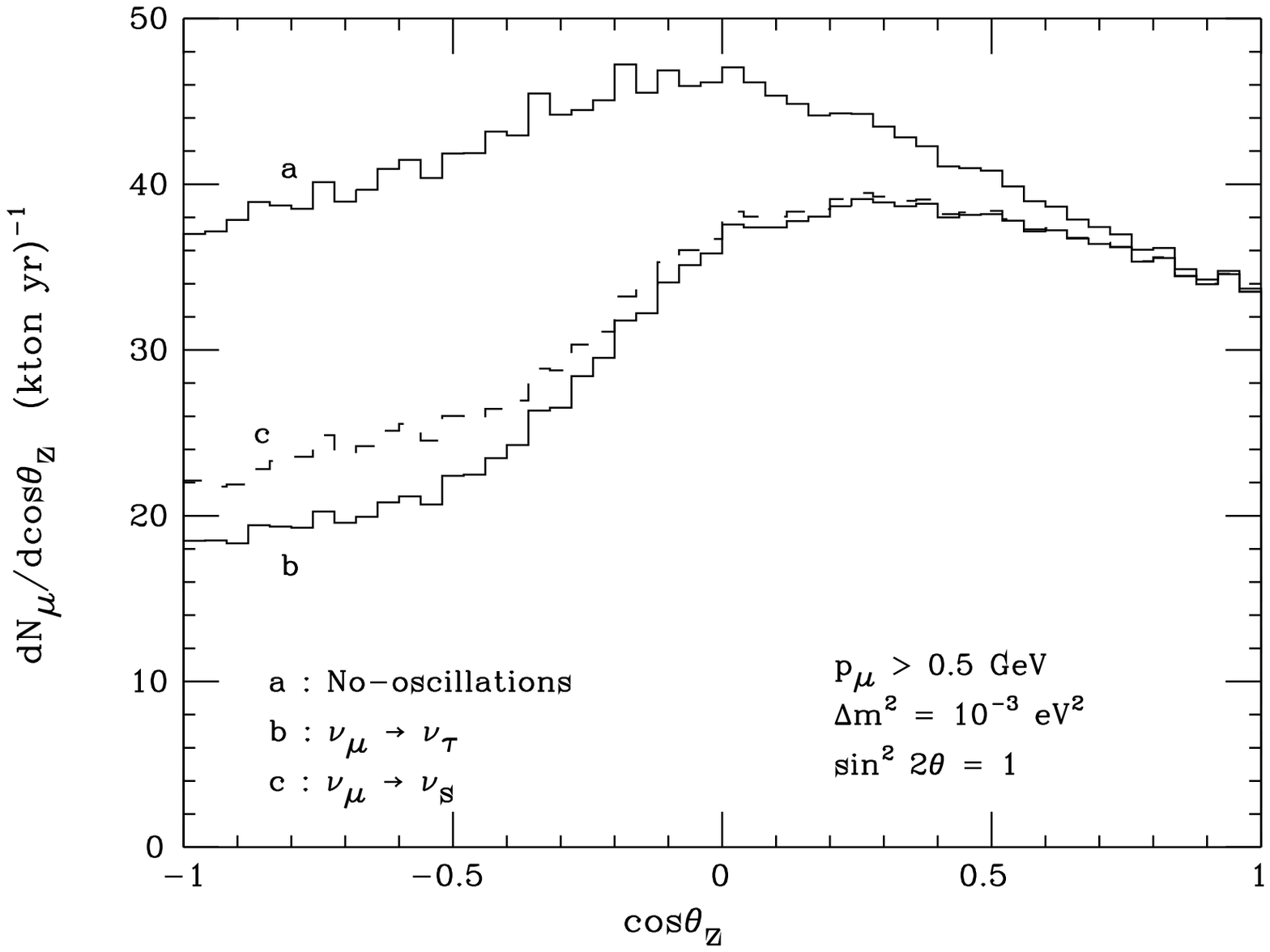,height=9cm}}
\vspace {0.4 cm}
\centerline{\psfig{figure=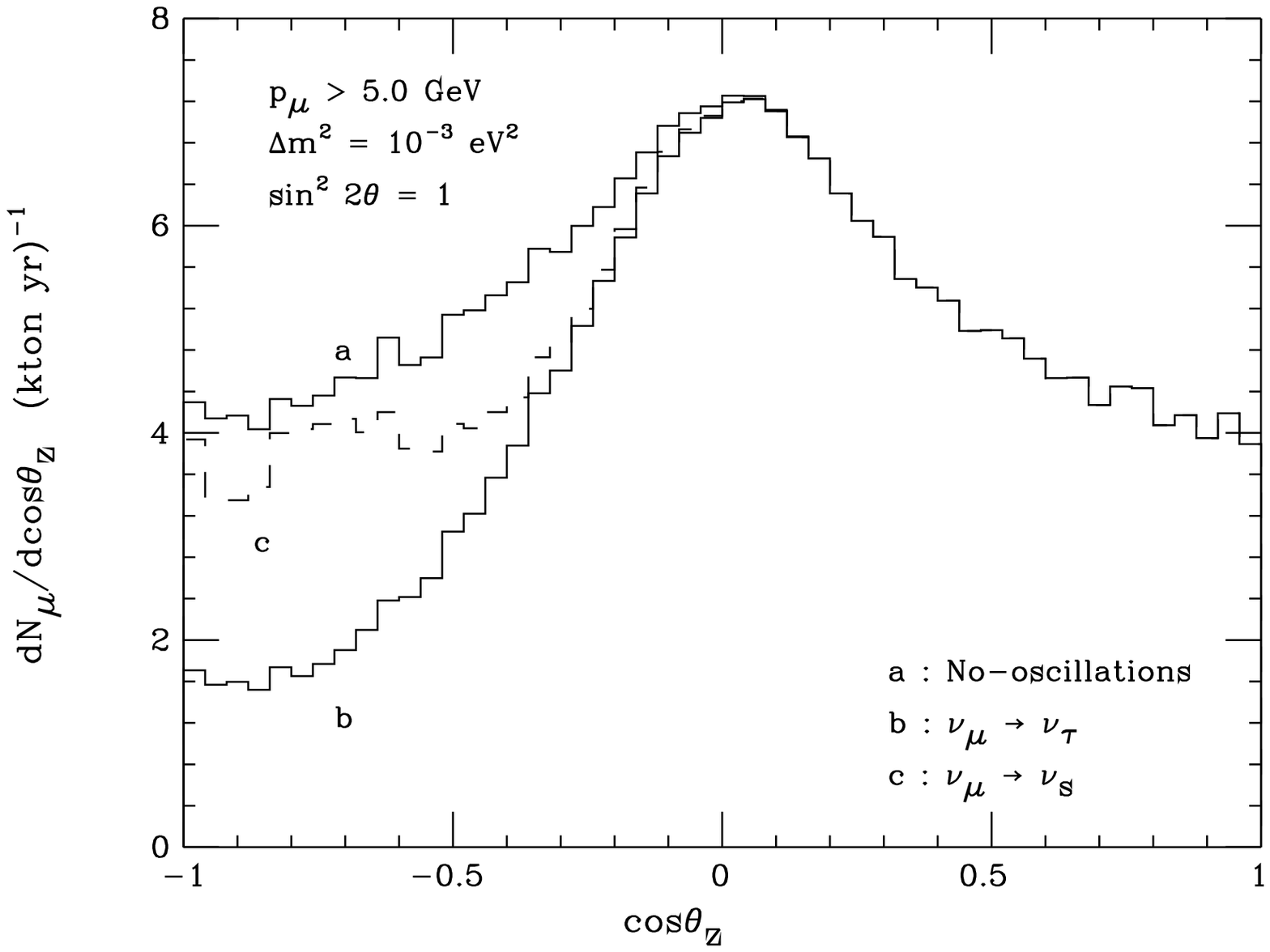,height=9cm}}
\vspace {0.4 cm}
\caption{Upper panel: Angular  distribution  of the  muon  events 
with $p_\mu \ge 0.5$~GeV, in the absence of oscillations (histogram a),
and in the presence of  oscillations   with  $\Delta m^2 = 10^{-3}$~eV$^2$
and  maximal $\nu_\mu$--$\nu_\tau$  (b)  or 
$\nu_\mu$--$\nu_s$  (c) mixing.
Lower panel: the same, with $p_\mu \ge 5.0$~GeV.
\label{fig_ang1} }
\end{figure}

\begin{figure} [t]
\centerline{\psfig{figure=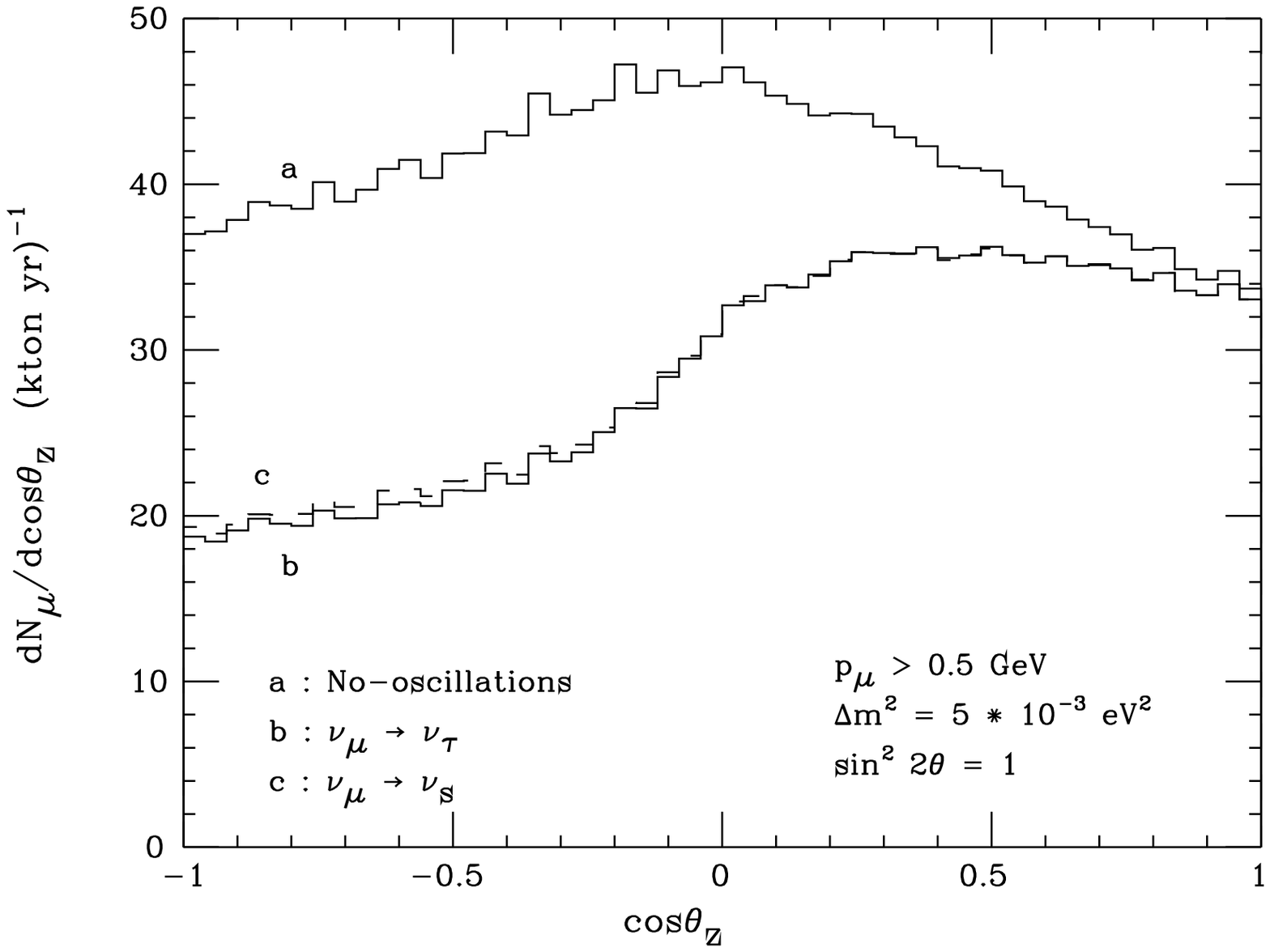,height=9cm}}
\vspace {0.4 cm}
\centerline{\psfig{figure=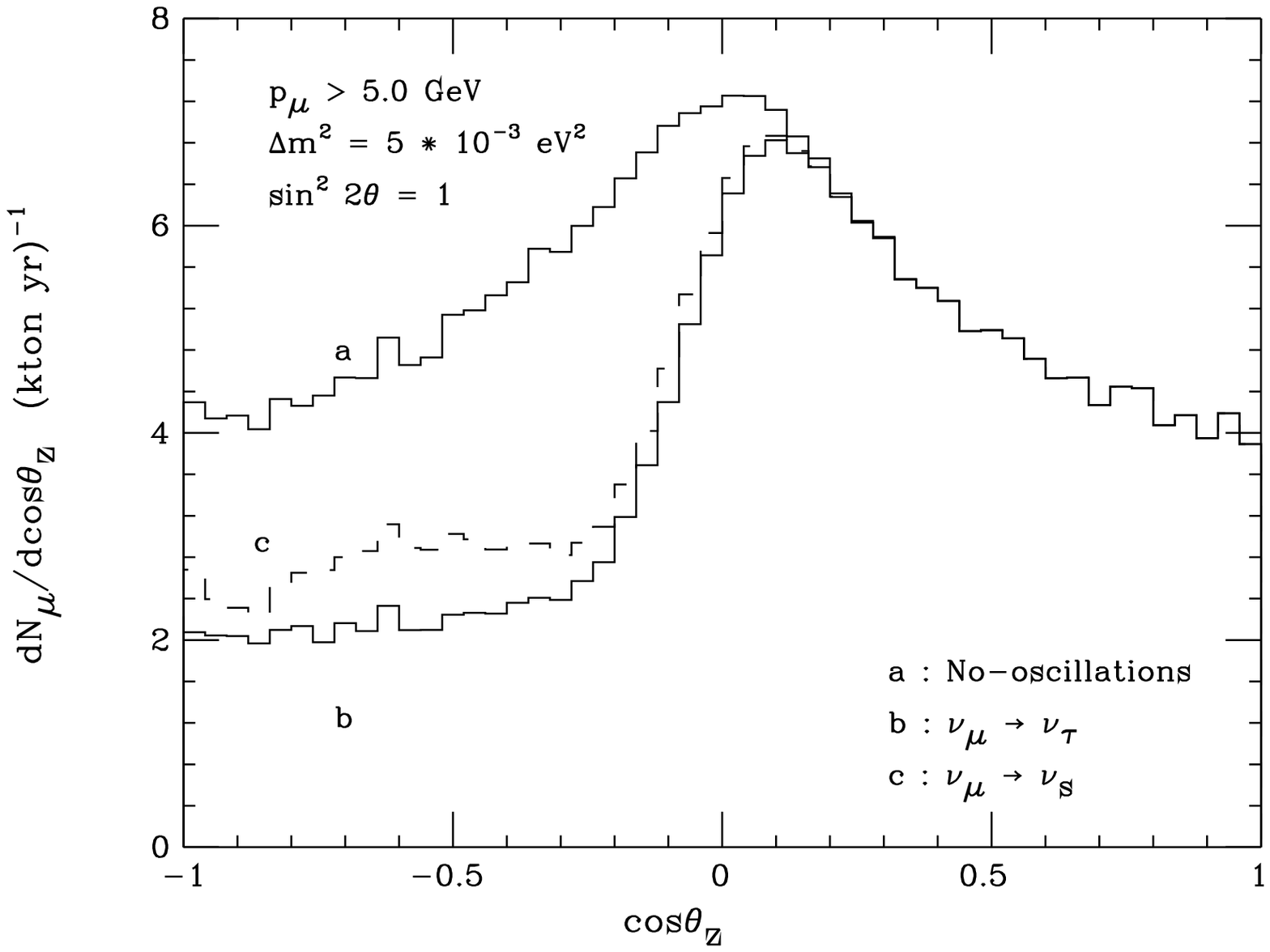,height=9cm}}
\vspace {0.4 cm}
\caption{As in fig.~\protect\ref{fig_ang1},   with
$\Delta m^2 = 5 \cdot 10^{-3}$~eV$^2$.
\label{fig_ang2} }
\end{figure}

As an illustration  in fig.~\ref{fig_ang1} and~\ref{fig_ang2}  we 
show the zenith angle distributions of  muons 
with momentum  $p_\mu \ge 0.5$~(5.0)~GeV  produced  inside   the detector
in charged  current neutrino interactions,  and the effects  produced
by  $\nu_\mu  \leftrightarrow \nu_\tau$ 
and $\nu_\mu  \leftrightarrow \nu_s$ oscillations
for  two  representative  choices of    oscillation parameters:
maximal  mixing and $\Delta m^2 = 10^{-3}$~eV$^2$ (fig.~\ref{fig_ang1})
or  $\Delta m^2 = 5\cdot 10^{-3}$~eV$^2$ (fig.~\ref{fig_ang2}).
In order to generate these  figures we have 
used  a Montecarlo code based on the neutrino flux
from Bartol \cite{Bartol} and  the neutrino cross section as  described in ref.
\cite{LLS};  to  compute the oscillation
probabilities we  have also 
generated  the neutrino production point according
to the   analytical  approximations  described  in ref.~\cite{nu-position}.
The  event  rates  calculated (for the position of the Gran Sasso 
underground laboratory) in the absence of oscillations
are  82  and 10  events/(kton~yr). No effects of detector efficiency  
and resolution have been included.

In the presence of
oscillations the   rate  of muons   events with up--going directions
($\cos \theta_z < 0$) are strongly depleted.  For the events  with 
$p_\mu \ge 0.5$  (upper  panels) the angle between the neutrino and the muon
has  a broad  distribution with average
 $\langle \theta_{\mu\nu} \rangle = 26^\circ$  and this explains
the loss of events   also
for downward--going directions.  
For the   higher  momentum events  ($p_\mu \ge 5$~GeV, lower  panels )
the muon  direction  is  more
tightly correlated with the parent neutrino ($\langle  \theta_{\mu\nu} 
\rangle =6.5^\circ$),  and  the depletion of  down--going events is 
much smaller.

The suppression of the  rate  in the case of  $\nu_\mu$--$\nu_s$ oscillations
is  smaller than in the $\nu_\mu$--$\nu_\tau$ case, because the
effective  mixing is  reduced  by matter  effects.
The difference   between the two cases  
depends  on the   muon momentum  considered  (that is   strongly correlated
with the parent  neutrino  energy).  For  low  muon momentum 
the oscillation   probabilities   of the two    hypotheses  are identical
or  very  similar.  With increasing    momentum the matter  effects  become
progressively more  significant   reducing  the $\nu_\mu \leftrightarrow \nu_s$
oscillation probability.
This  different energy  dependence  of the  oscillation
probabilities is  in fact the  most  easily  detectable  signature of
the existence of mixing with sterile neutrinos.

The effect is  also   strongly  dependent  on the value of
$\Delta m^2$. In the case $\Delta m^2 = 10^{-3}$~eV$^2$ (fig.~\ref{fig_ang1})
the suppression 
$N_\mu/N_\mu^\circ$    of  the muon rate with respect to the
no-oscillation prediction,   for 
the   events with  $p_\mu \ge 0.5$~GeV is 0.74 (0.78)
in the cases of $\nu_\mu$--$\nu_\tau$  ($\nu_\mu$--$\nu_s$) mixing;  for 
the higher   momentum cut the suppression 
becomes  0.83 (0.94)  for the two  cases, with a much  larger  difference.
The angular  dependence  of the 
suppression is also different   
for oscillations into tau or sterile neutrinos \cite{footnote-angular}.

For  the higher  value $\Delta m^2 = 5 \cdot 10^{-3}$~eV$^2$ 
(fig.~\ref{fig_ang2})
the   suppression is    nearly    identical    
(0.689 and  0.695)  in the case of  the  lower  momentum  muons
and   also for the higher  momentum sample
the difference between the two cases 
with suppressions  of 0.74  and 0.78 is small.

\begin{figure} [t]
\centerline{\psfig{figure=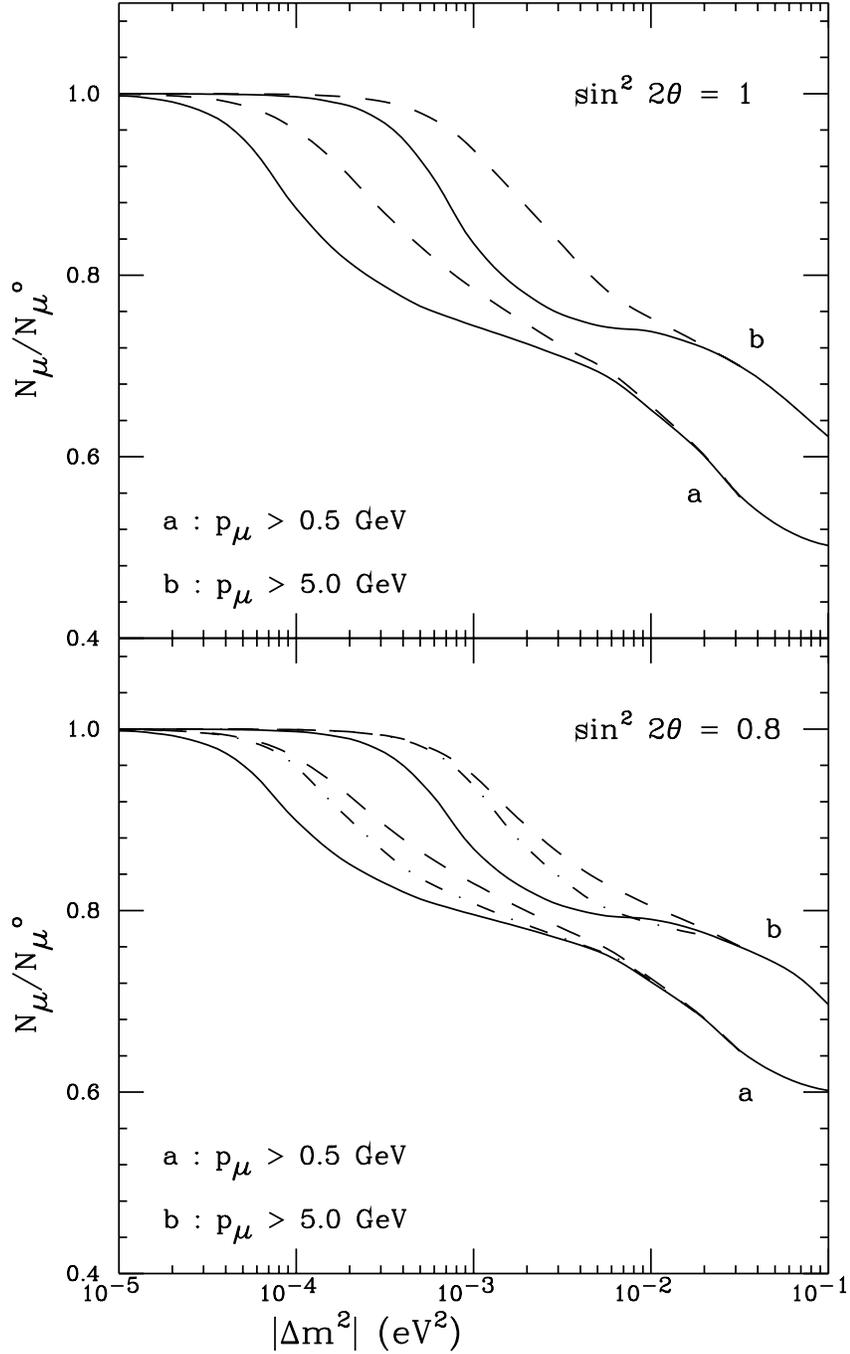,height=18cm}}
\vspace {0.4 cm}
\caption{Ratio of the rate of muon events to the no--oscillation case
as a function of $\Delta m^2$, for two  choices of the minimum
muon momentum.
The upper (lower)  panel   is  calculated for
$\sin^2 2 \theta = 1$ (0.8). The solid (dashed) curves  are  for
$\nu_\mu$--$\nu_\tau$  
($\nu_\mu$--$\nu_s$) mixing. In the lower panel the
dot--dashed  curve describes 
$\nu_\mu$--$\nu_s$    mixing  with $\Delta m^2 <0$. 
\label{fig_rr} }
\end{figure}

To summarize this  discussion in fig.~\ref{fig_rr} we plot 
the  ratio $N_\mu/N_\mu^\circ$ of the  muon rate to
the   prediction in the absence of oscillations, 
as a function of $\Delta m^2$
for the two different cuts in the muon momentum, $p_\mu \ge 0.5$~GeV or
$p_\mu \ge 5.0$~GeV, both for maximal mixing (upper half of the figure) and
for $\sin^2 2\,\theta$ = 0.80 (lower half). The predictions for  
$\nu_\mu \leftrightarrow \nu_\tau$ oscillations are given by the full lines,
those for oscillations into sterile neutrinos by dashed (dot--dashed) 
lines  assuming  $\Delta m^2 >0$  ($\Delta m^2 < 0$).
For very small  $|\Delta m^2|$   the oscillation cannot
develop and  the  ratio $N_\mu/N_\mu^\circ \simeq 1$; with increasing  
$|\Delta m^2|$  the ratio   decreases    monotonously,
reaching  asymptotically a   value 
$1 -{1\over 2} \sin^2 \,2 \theta$,  corresponding to the averaging
over rapid  oscillations. Note that the region
$|\Delta m^2| \sim 2$--$4 \cdot 10^{-3}$~eV$^2$  corresponds 
to  a situation where most  up--going  (down--going) neutrinos do (do not)
oscillate.  From fig.~\ref{fig_rr}   we can  see that for
sufficiently large  $|\Delta m^2|$  the  curves  corresponding to the
$\nu_\mu$--$\nu_\tau$  and  $\nu_\mu$--$\nu_s$   mixing  become
undistinguishable.   Selecting  higher   momentum   muons
the  difference     between    the two oscillation
hypotheses    remains    significant until larger
values of  $|\Delta m^2|$,  however the   event  sample  available to study
the effects  becomes  progressively smaller.

A larger  number of  high energy neutrino events 
is  obtainable  from    the upward--going  muon sample.
In the  remaining sections  we  discuss this  class  of  events.

\section {Upward--going muons}

Muon  neutrinos  and  antineutrinos
can  be  detected indirectly observing the
muons  produced  in CC interactions
in the vicinity of the detector.
In the presence of  $\nu_\mu \to \nu_x$ 
oscillations the flux of $\nu$--induced muons is
suppressed,  the energy spectrum  is deformed 
and the angular  distribution distorted.
In essentially  the entire region  of 
$\nu$--oscillations parameter space that  is  a solution
for  the   atmospheric  neutrino problem, the muon
flux is modified  in ways  that are in principle observable
with large area  detectors. The neutrino  energies that
contribute to the  muon flux  extend  up to
$\sim 1000$~GeV, therefore the matter effects are important 
and  allow in principle to distinguish between
the  $\nu_\mu$--$\nu_\tau$  and $\nu_\mu$--$\nu_s$ hypotheses.

The  $\nu$--induced muon flux is  measurable 
only in directions from below the horizon
\cite{nu-induced-muons}, therefore it is not possible to measure
deviations from the up--down symmetry of 
the no--oscillation flux. The 
effects  of oscillations can be  studied only 
comparing the data  with a  Montecarlo calculation, based on
assumptions about the primary cosmic ray  flux,  the properties
of particle production in hadronic interactions  with air--nuclei, 
the charged  current neutrino cross section, and  the 
properties of muon propagation.   The theoretical  uncertainties in
the  different elements  of the   calculation
result in a sytematic  uncertainty 
of $\sim 20\%$ for the  absolute normalization of the flux.
The energy  and  especially the 
zenith angle dependence  of the   flux
can be  calculated  more reliably \cite{Lipari-Lusignoli}  
and   are  useful  probes of  the existence of oscillations.

\begin{figure} [t]
\centerline{\psfig{figure=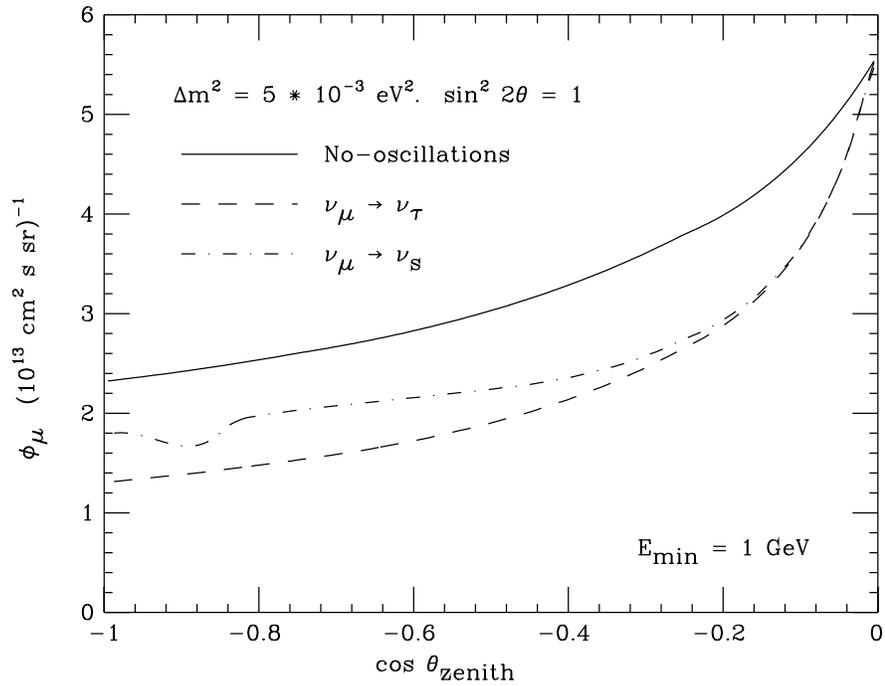,height=9cm}}
\vspace {0.4 cm}
\caption{Upward--going muon flux as a function of zenith angle (with
$E_{\rm min} = 1$~GeV), in the absence of oscillations
(solid line), and  for  maximal mixing 
and  $\Delta m^2 = 5 \cdot 10^{-3}$~eV$^2$
in the cases  of $\nu_\mu \to \nu_\tau$ (dashes) and
$\nu_\mu \to \nu_s$ (dot--dashes). 
\label{fig_upmu0} }
\end{figure}

\begin{figure} [t]
\centerline{\psfig{figure=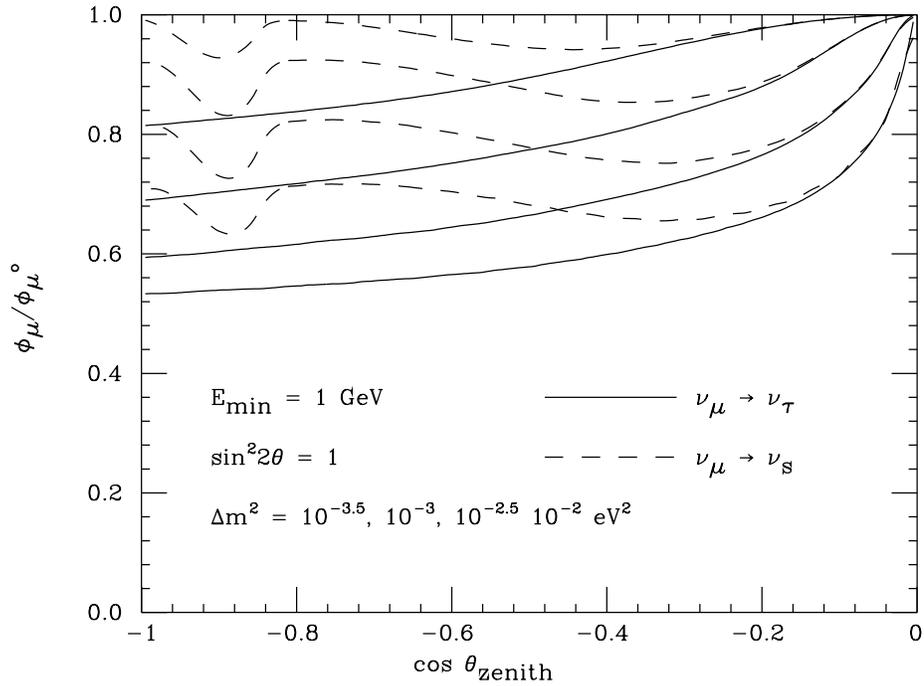,height=9cm}}
\vspace {0.4 cm}
\caption{Ratio between the upward--going muon flux ($E_\mu \ge 1$~GeV),
and the no--oscillation  prediction for  maximal  mixing and 
(starting from the highest line) $\Delta m^2 = 10^{-3.5}$,
$10^{-3}$, $10^{-2.5}$ and $10^{-2}$~eV$^2$. The solid
(dashed) lines are for $\nu_\mu$--$\nu_\tau$ ($\nu_\mu$--$\nu_s$) mixing.
\label{fig_upmu1} }
\end{figure}

As an illustration of the effect of neutrino oscillations,
in fig.~\ref{fig_upmu0} we show  the flux  of 
upward going muons  above a minimum energy
$E_{\rm min} = 1$~GeV  plotted as  a  function of the  zenith angle.
The flux is
calculated in the absence of oscillations, and for
$\nu_\mu$--$\nu_\tau$ and 
$\nu_\mu$--$\nu_s$   oscillation parameters corresponding to
the SK `test point' 
($\Delta m^2 = 5\cdot 10^{-3}$~eV$^2$ and  $\sin^2 2 \theta= 1$).
As a best estimate of the muon flux  we have used the 
neutrino flux  of Bartol \cite {Bartol}, the cross
sections  described in \cite{LLS} and the muon
energy loss of \cite{Lohmann}.
The no--oscillation  distribution has a maximum  (minimum) for 
horizontal (vertical)  muons,  that reflects the larger flux  of 
neutrinos  for  inclined    directions.  
In reference \cite{Lipari-Lusignoli} we  have  discussed in some detail
the prediction of the upgoing muons zenith angle
distribution and  its theoretical uncertainties.
Here we  will concentrate on the shape of the distortion of the
distribution produced by oscillations.
In fig.~\ref{fig_upmu1}, we have 
plotted the 
ratio between the oscillated  and non--oscillated  flux
 for different  values of  $\Delta m^2$, always 
assuming  maximal  mixing.
In the case of $\nu_\mu \leftrightarrow \nu_\tau$ oscillations the
distortion factor  varies    smoothly and  monotonously  from a
value close to  unity for horizontal  muons, to  a maximum
suppression   for vertical  muons.
This behaviour  can be easily  understood:
the neutrinos  that generate the  upward--going muons
have a broad  energy  spectrum,    with a  shape  that  
(in the absence of oscillations) changes
only moderately for   different  zenith angles.
For a  fixed value  of the zenith angle,
the transition  probability   $P(\nu_\mu \to \nu_\tau)$
is  a function of the  neutrino  energy, 
for low  $E_\nu$   it  oscillates rapidly  
with  an average value
${1\over 2}\sin^2 2 \theta$, for large  $E_\nu$ it   decreases as
$L^2/E_\nu^2$,  and  is  negligibly  small  for 
$E_\nu \gg L\, |\Delta m^2| \simeq 2 R_\oplus\, |\cos \theta_z|\,|\Delta m^2|$.
With increasing $|\cos \theta_z|$ neutrinos of higher energy have a chance 
to  oscillate and  the muon flux is  more  suppressed.

In the case of oscillations  with maximal mixing into  sterile neutrinos,
the suppression of the flux is smaller  than
in the corresponding vacuum  case  because  
the effective  mixing in matter  is  reduced. The distortion 
factor does not  depend monotonously on the zenith angle,
but presents a structure with 
two  minima at $\cos\theta_z \simeq -0.35$ and $-0.9$.
The origin of this structure  can be understood as  a simple
consequence of the  considerations  developed   in the  
discussion of oscillation probabilities.
The  transition  probability $P(\nu_\mu \to \nu_s)$
as  a function of energy for  a  fixed  zenith angle,
is  similar to the  vacuum case for  low  $E_\nu$.
For  larger $E_\nu$ it assumes   the   form  discussed
in section 2.2,    developing  a characteristic  shape
(see  the lower panel  in fig.~\ref{fig_prob_a}) that  after  integration
over all neutrino energies remains visible in  fig.~\ref{fig_upmu1}.

The  detection of the structure in the
distortion factor, present at
$\cos\theta_z \simeq -0.9$ would be a 
clear and  unambiguous experimental proof of the existence  of 
mixing  between muon and sterile neutrinos,  but 
requires  a large number of events. 
Easier to detect is  perhaps the 
general  trend  of a suppression with respect to the 
no--oscillation expectation 
that becomes  weaker   going from $\cos \theta_z \simeq -0.3$
to $\cos \theta_z \simeq -0.8$, in constrast with the vacuum case.

\begin{figure} [t]
\centerline{\psfig{figure=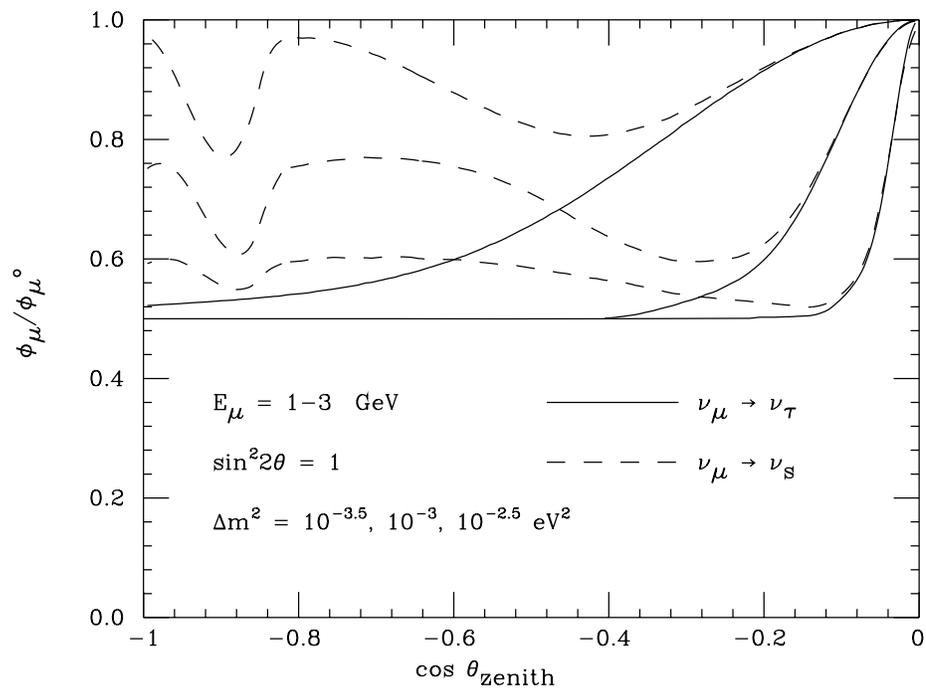,height=9cm}}
\vspace {0.4 cm}
\caption{As in fig.~\protect\ref{fig_upmu1}, with the upward going muon flux
calculated in the  energy interval $1 \le E_\mu \le 3$~GeV.
\label{fig_upmu2}  }
\end{figure}

The qualitative features of the  distortion  factors described  above
are present also selecting   other energy intervals for
the uwpard going muons.
In fig.~\ref{fig_upmu2}  we  show the
distortion factor   of the zenith  angle  distribution
of upgoing muons  in the  interval $1 \le E_\mu \le 3$~GeV
selected because it is in principle experimentally accessible,
the muon  energy being estimated from the measured  range for
stopping particles.
The interval  of parent neutrino
energies that  contributes to this   flux 
is narrower, extending up to 
a  neutrino energy of $\sim  50$~GeV.
This  can  result  in  strong  features in the angular  distribution
for $\Delta m^2$ smaller that $\sim 10^{-3}$~eV$^2$; 
on the other hand  for $\Delta m^2$ approaching $10^{-2}$~eV$^2$
the   distortion factor rapidly reaches the 
constant value $ 1 - {1\over 2} \sin^2 2\theta$ both for vacuum and for 
matter oscillations.

\begin{figure} [t]
\centerline{\psfig{figure=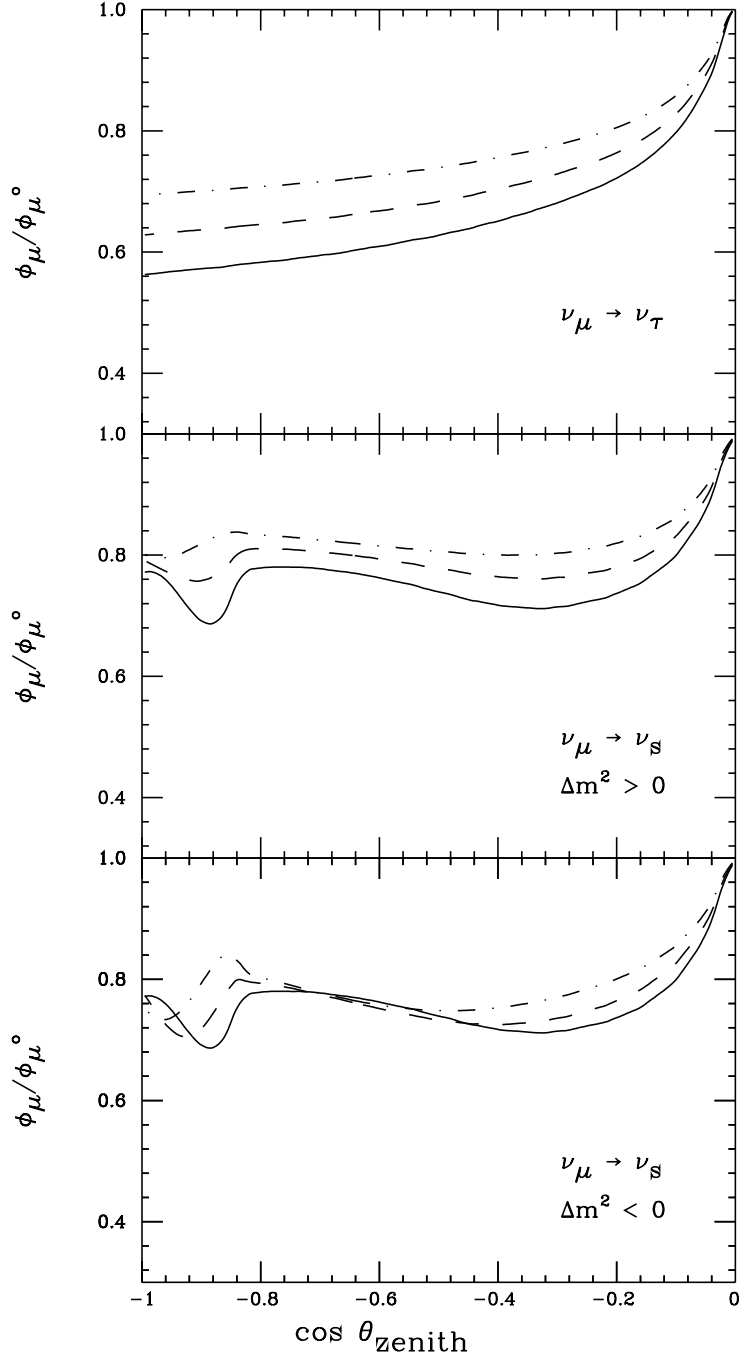,height=18cm}}
\vspace {0.4 cm}
\caption{Ratio between the upward--going muon flux ($E_\mu \ge 1$~GeV),
and the no--oscillation  prediction for  
$|\Delta m^2| = 5 \cdot 10^{-3}$~eV$^2$ and  $\sin^2 2 \theta = 1$
(full curves), 0.9 (dashed) and 0.8 (dot-dashed). 
The upper panel is for $\nu_\mu$--$\nu_\tau$ mixing, the middle
(lower)  panel  for $\nu_\mu$--$\nu_s$ mixing  and
positive (negative) $\Delta m^2$.
\label{fig_upmu3}  }
\end{figure}

In fig.~\ref{fig_upmu3}    
we show  the distortion factors   of  the flux of  muons with
$E_\mu \ge 1$~GeV for $\Delta m^2 = 5\cdot 10^{-3}$~eV$^2$  and
$\sin^2 2 \theta = 0.7$.
In the  case of  vacuum oscillations the distortion factor
for $\sin^2 2\theta < 1$ can be obtained from the
result for maximal mixing  simply as:
\begin{equation}
\langle P(\Delta m^2, \sin^2 2\theta)\rangle = \sin^2 2\theta\, 
\langle P(\Delta m^2, 1) \rangle 
\end{equation}
where
$\langle P\rangle = 1 - \phi_\mu/\phi_\mu^\circ$  is the 
fractional  difference  between the non oscillated  and oscillated
flux.
In the presence of matter  effects such a simple  relation does not
exist,  and the dependence   on the mixing angle in
non trivial. 
The  structure in the   distortion factor   is in general less
visible in the case of non--maximal  mixing. This is  because the 
oscillation lengths  of neutrinos  and antineutrinos  are
different in the crucial  region around  the 
MSW resonance  (see  the  curves labeled  `c' and `d'  
in fig.~\ref{fig_osc2})  and this results in  different  angular patterns
of the distortion for neutrinos and antineutrinos,
as shown in  fig.~\ref{fig_prob_b}.  It is important also  to note that the
sign of the mass  difference, non observable in vacuum, is  important 
in this case.

\section {Upward--going muon data}    
Several   detectors have collected  data on neutrino--induced muons:
Baksan \cite{Baksan-upmu},
MACRO \cite{MACRO-upmu}, Kamiokande \cite{Kam-upmu},
IMB \cite{IMB-upmu}, and Super--Kamiokande  \cite{SK}.
The situation at present is  somewhat confused  because of  the poor
agreement  between  different experiments, and   between the data
and  theoretical  expectations  even after the inclusion of 
oscillations.

To compare theoretical predictions with experimental data we 
assume that the theoretical estimate  of the flux  is affected by 
systematic  uncertainties that  to a  good approximation
enter as a constant  normalization  factor  without affecting
the shape of the distribution \cite {Lipari-Lusignoli}.
We therefore define a $\chi^2$ as:
\begin{equation}
\chi^2 = {\rm min}_\alpha \left \{
\sum_j \left ( { \phi_j - \alpha \; \phi_j^{th} \over \Delta\phi_j} \right)^2 
+ \left ( {1 -\alpha \over \Delta \alpha } \right )^2 
\right \}
\label {eq:chi2}
\end{equation}
In eq.~(\ref{eq:chi2}),
the sum is  over the number of bins in  zenith  angle  (all experiments
have been using  10  bins of equal  width $\Delta \cos \theta_z = 0.1$),
$\phi_j$ and $\Delta \phi_j$ are  the  measured  value of the  flux  in
the bin and its error, 
$\phi_j^{th}$   is  the calculated  value  and 
$\alpha$ is  a  normalization  factor and
$\Delta  \alpha $
is the theoretical error, estimated as $\sim 0.2$.
The  combination of  experimental   errors  ($\Delta \phi_j$'s) and 
theoretical  uncertainty ($\Delta \alpha$) in the estimate  of the 
$\chi^2$ in eq.~(\ref{eq:chi2}) cannot  be  rigorously 
justified, but at  least qualitatively has  the  desired  features.
Note that  the limit  $\Delta \alpha \to 0$  forces $\alpha$ to be 1,
and  corresponds to   neglecting  the theoretical  uncertainties; 
the  opposite  limit  $\Delta \alpha \to \infty$  takes 
into consideration only the shape of the zenith angle  distribution.

\subsection {Super--Kamiokande}
The Super--Kamiokande collaboration \cite{SK} has  presented
data on  through--going muons, that is
tracks that  enter  in  and  exit from the  detector, with a minimum
path--length $L_{\rm min} = 7$~m.
To compare with the experimental data  we have  calculated
a muon flux as:
\begin{equation}
\phi^{th} (\cos \theta_z) =
{1 \over A(L_{\rm min}, \theta_z) } ~
\int_{E(L_{\rm min})}^\infty dE_\mu~ {d\phi_\mu^{th} \over dE_\mu}
 (E_\mu, \cos\theta_z) [A(L_{\rm min}),\theta_z) - A(L(E_\mu),\theta_z)]
\label {eq:flux-SK}
\end{equation}
where 
$d\phi^{th}_\mu/dE_\mu$ is the   flux  of neutrino  induced muons
with  energy $E_\mu$ and zenith angle $\theta_z$,
$L(E)$ and  $E(L)$ are the muon range in water and its inverse  function,
and $A(L, \theta_z)$ is  the projected area  of the detector
that  corresponds  to  trajectories with internal pathlength longer than $L$.
The factor
$[A(L_{\rm min},\theta_z) - A(L(E_\mu), \theta_z)]$ weights the 
contribution of  muons  of energy $E_\mu$ to the  flux, taking into 
account the fact that muons  of low  energy  stopping 
in the detector should  not  be  counted.
Assuming  for  Super--Kamiokande the ideal  geometry of a cylinder
(with radius $R = 16.9$~m and  height $H = 36.2$~m), a simple geometrical 
calculation gives:
\begin{equation}
A(L, \theta_z) =
  2\,R\,H\,\sin\theta_z ~\sqrt {1 - x^2}
   +  2\,R^2 |\cos \theta_z| ~[ \cos^{-1} x 
   - 3\,x \,\sqrt {1 - x^2} ] \; \Theta \bigl[L_{max}(\theta_z) - L \bigr]\;,
\label{eq:acceptance}
\end{equation}
with $x = L\sin\theta_z/2R$ and
$L_{\max}(\theta_z) = {\min }\,[2\,R/\sin\theta_z$, $H/|\cos\theta_z|]$.

Our calculation of the $\nu$--induced muon flux  is in excellent agreement 
with the 
one  presented by the SK  collaboration \cite{SK}, indicating  that our  
simple  geometrical picture of the acceptance  is a  reasonable approximation.
The main  difference   between our    results  
and the theoretical curves  presented by SK 
 is a  higher  ($\sim  5\%$)  absolute normalization 
of our  estimate, probably due  to the fact  that we model the neutrino 
cross section
including an explicit  treatment  of  the quasi--elastic scattering and
single--pion   production \cite{LLS}. 

\begin{figure} [t]
\centerline{\psfig{figure=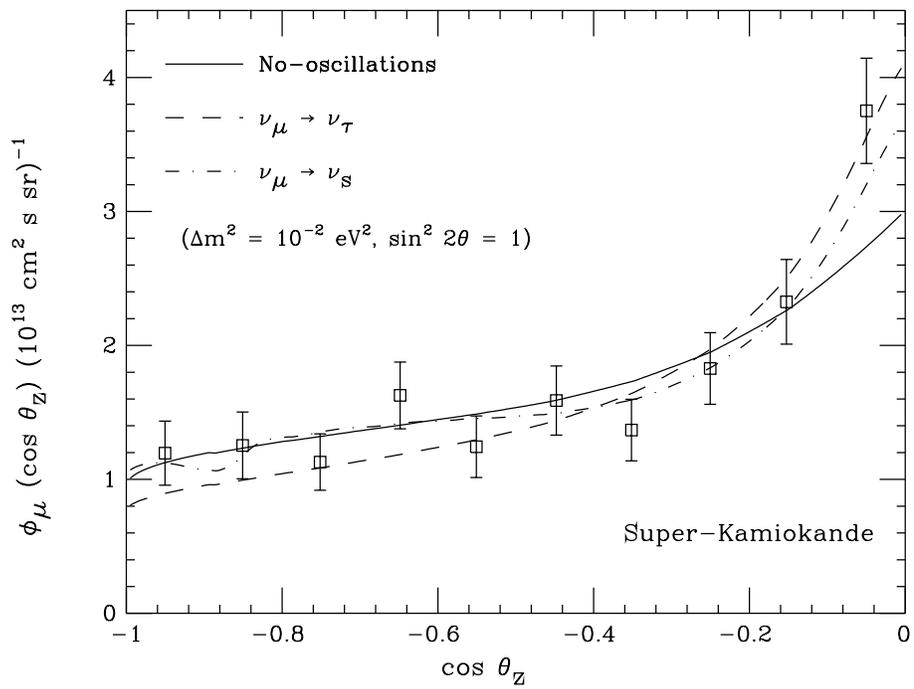,height=9cm}}
\vspace {0.4 cm}
\caption{Flux of through--going upward going muons  
measured  by Super--Kamiokande
\protect\cite{SK},  compared  with  theoretical predictions  with and
without oscillations (see text)
\label{fig:SK} }
\end{figure}

An illustration of the effect of  oscillations  
is  shown in fig.~\ref{fig:SK}, where 
we compare the SK data  points  with 
(i) a no--oscillation  prediction (full line),  (ii) 
$\nu_\mu \leftrightarrow \nu_\tau$ oscillations  with 
$\sin^2 2 \theta = 1$ and
$\Delta m^2= 10^{-2}$~eV$^2$  (dashed line), (iii)
$\nu_\mu \leftrightarrow \nu_s$ oscillations  with the same  parameters  
(dot--dashed line).  
The three  models  have  been  calculated   with the same  theoretical  input
($\phi_\nu$ from  Honda et al. \cite {HKKM}, $\sigma_\nu$ as described in
\cite {LLS} with   leading order PDF's from
GRV \cite{GRV}), and rescaled with factors $\alpha = 0.88$, 1.21 and
1.09.
All three models  give  acceptable  fits to
the data, with $\chi^2$ values of 12.8, 10.0  and 6.2.
We can observe: (i) the  no--oscillation  hypothesis
tends  to be too flat,  missing the data point for  near--to--horizontal muons; 
(ii) the  hypothesis of $\nu_\mu \leftrightarrow \nu_\tau$ oscillations  with  
maximal   mixing gives  a  high ratio  between the horizontal  and  
vertical  flux, and in this hypothesis the best fit ($\chi^2 \simeq 8$)
is obtained for $\sin^2 2 \theta \simeq  0.70$;
(iii) the   hypothesis  of $\nu_\mu \leftrightarrow \nu_s$  oscillations with 
maximal  mixing gives a suprisingly  good  fit  to the data.
We would like to stress the importance of the measurement   
of near--to--horizontal  muons ($-0.1<\cos \theta_z <0$) 
in  determining  the results of the  fits.

For the no--oscillation hypothesis, 
the  $\chi^2=12.8$   has  been  computed  assuming  $\Delta \alpha = 0.2$.
Assuming  instead $\Delta \alpha = 0$   the $\chi^2$  takes the  value
20.4  (probability $2.6\cdot 10^{-2}$).    Neglecting 
the absolute normalization (assuming $\Delta \alpha \to \infty$)
the optimum  $\alpha$   becomes   0.87   with $\chi^2$ decreasing to 12.4;
these numbers  can  be compared  with  the estimates: 
 $\alpha = 0.83$ and $\chi^2 = 12.7$ 
of the SK   collaboration \cite {SK}
done with  the same   neutrino flux \cite {HKKM}.
Using the Bartol  neutrino flux \cite {Bartol} we find a
 result that is 6.5\%  higher  in absolute  normalization
with small differences  in the angular  distribution: 
for $\Delta \alpha = 0.2$
the $\chi^2$  for the no--oscillation  hypothesis  is 15.1.

\begin{figure} [t]
\centerline{\psfig{figure=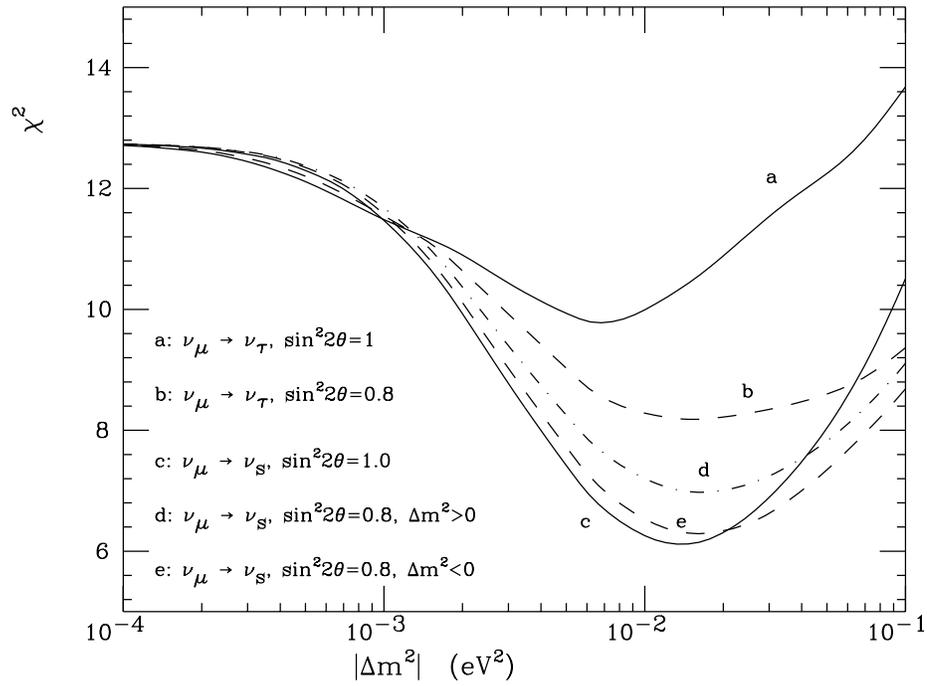,height=9cm}}
\vspace {0.4 cm}
\caption{$\chi^2$  of  fits to the upward--going muons  data 
of Super--Kamiokande calculated according to eq.~\protect\ref{eq:chi2}
plotted as a function of $\Delta m^2$, for  $\sin^2\,2 \theta = 1$ and 0.8
assuming $\nu_\mu$--$\nu_\tau$ and
$\nu_\mu$--$\nu_s$ mixing.  The uncertainty $\Delta \alpha$  on the 
absolute normalization is  0.2.
\label{fig:SK1}
}
\end{figure}

The $\chi^2$ resulting from fits 
to the data using the oscillation hypotheses (either 
$\nu_\mu \leftrightarrow \nu_\tau$
or  $\nu_\mu \leftrightarrow \nu_s$) are presented in fig.~\ref{fig:SK1}
as a function of $\Delta m^2$ for    $\sin^2\,2\theta=1$ and~0.8. 
It is  interesting  to note that the  inclusion of neutrino oscillations,
with    parameters  in the  range suggested  by the analysis of 
(semi--)contained events, 
does  result in  a better
agreeement  between  data and  prediction.
The best fit  ($\chi^2 \simeq 6.1$)  is  
obtained for $\nu_\mu \leftrightarrow \nu_s$ oscillations
with $\Delta m^2 \simeq  1.4\cdot  10^{-2}$~eV$^2$ 
and $\sin^2 2\,\theta \simeq 1$ (with $\alpha \simeq 1.12$).
Assuming $\nu_\mu \leftrightarrow \nu_\tau$ oscillations 
the $\chi^2$ is minimized ($\chi^2 \simeq 7.8$ )
for  $\Delta m^2 \simeq 2.2 \cdot 10^{-2}$~eV$^2$ and 
$\sin^2\,2\theta \simeq 0.7$ (with $\alpha = 1.18$).

\begin{figure} [t]
\centerline{\psfig{figure=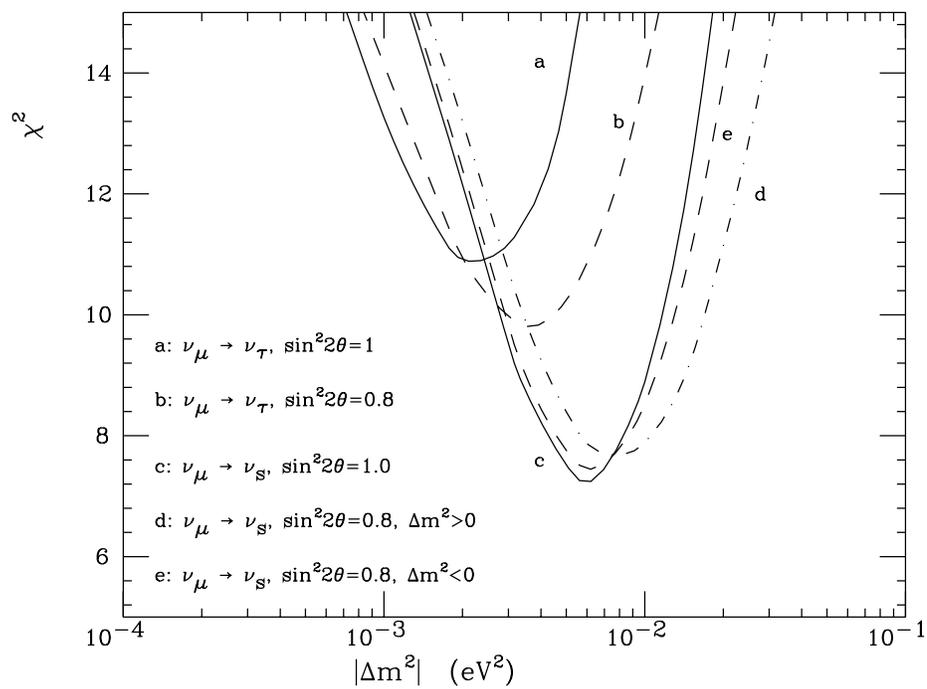,height=9cm}}
\vspace {0.4 cm}
\caption{$\chi^2$  of fits to the SK upgoing muons
(as in fig.~\protect\ref{fig:SK1})
calculated   with a fixed normalization ($\alpha =1$).
\label{fig:SK2}
 }
\end{figure}

Since our  estimate  of the theoretical  error $\Delta \alpha \simeq 0.2$ 
is  rather arbitrary, we have repeated the fits  for other 
choices of $\Delta \alpha$. The  $\nu_\mu \to \nu_s$ hypothesis  has always 
a lower $\chi^2$, because   it  predicts a  shape  that is 
in closer  agreement with the  data  and requires a  normalization factor
$\alpha$ closer to unity. Large values of $\Delta m^2$ require a normalization
factor considerably larger than unity, therefore 
 an increased  weight of the absolute 
normalization (i.e. smaller values  of $\Delta \alpha$)
lowers the best--fit values of $\Delta m^2$. The limiting case  
of $\Delta \alpha = 0$  is  illustrated in fig.~\ref{fig:SK2}.
It may be noted that in this  case,   the range of parameters
that minimize the  $\chi^2$ for   oscillations  into sterile neutrinos
is  very  similar to the one   that fits the sub--GeV 
and multi--GeV SK data  according to Foot et al. ,
as one can see comparing fig.~\ref{fig:SK2} with fig.~12 in 
\cite{Foot_etal}.

In summary, the SK  data on upward going  muons
are  consistent  with the existence
of  neutrino oscillations,  even if still inconclusive if  taken
by themselves. 
Considering the reduction of statistical errors 
expected  with  a longer  exposure,   future data 
should allow  to  distinguish the hypothesis
of $\nu_\mu$--$\nu_\tau$ and $\nu_\mu$--$\nu_s$  mixing.

\begin{figure} [t]
\centerline{\psfig{figure=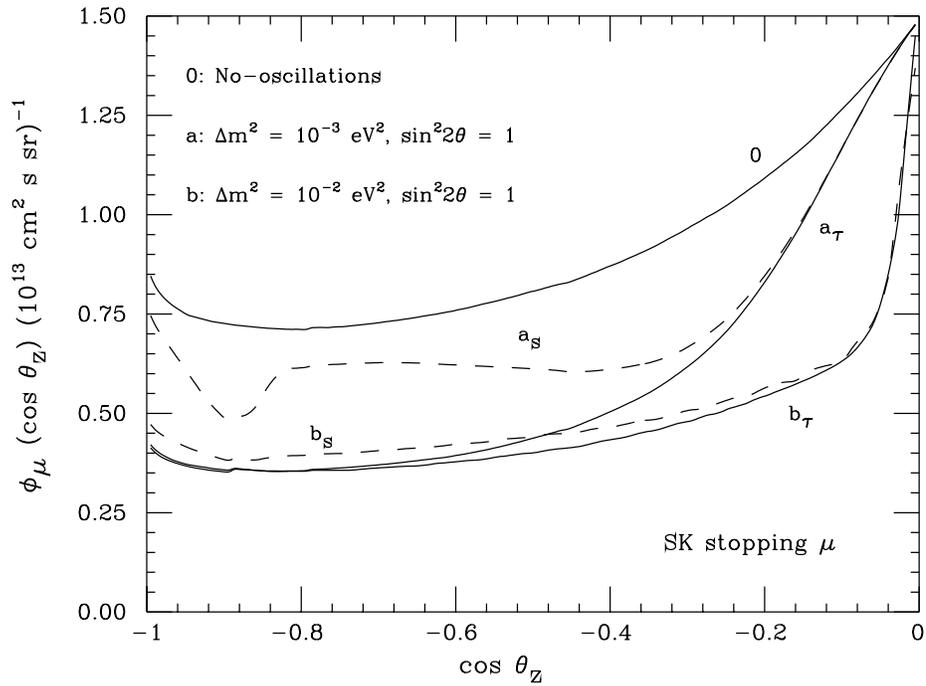,height=9cm}}
\vspace {0.4 cm}
\caption{Predictions of the rate  and angular  distribution of 
stopping muons  in SK,  in the absence  (and presence) of oscillations.
The subscript $\tau$  (s) indicates 
$\nu_\mu$--$\nu_\tau$   
($\nu_\mu$--$\nu_s$) mixing.
\label{fig:stop}
}
\end{figure}

Additional  information on the existence of  neutrino oscillations
can be obtained  from the analysis of  neutrino induced  muons
that stop in the  detector \cite{IMB-upmu}
(see also \cite{ALL93,Frati93,Lipari-Lusignoli} for  additional discussion).
In fig.~\ref{fig:stop}  we present  predictions of the absolute  
rates  and  angular  distributions  of upgoing muons stopping 
in Super--Kamiokande. As  discussed   in previous works, 
in the presence of  oscillations the  flux   of stopping  muons
is  suppressed  more strongly  than the  flux of  through--going particles.
The  typical  neutrino  energies
that  produce  stopping  muons  are similar  to  those that are the origin
of multi--GeV  events, and therefore matter  effects have   approximately
the same  importance.
The  suppression of the stopping flux and 
the distortion  of the angular distribution
depend  on the  type of oscillations only for  small $|\Delta m^2|$.

\subsection {Baksan and MACRO.}
Data samples on upgoing muons with comparable statistics 
\cite{note_area}
have also been collected by other
experiments.  
These data are difficult to describe
both with and without oscillations, and also in poor agreement 
with each other \cite{Fogli-upmu,SK}.

The Baksan  and MACRO experiment  have presented  their results
giving the  muon flux   above a threshold energy of  1~GeV
correcting  for the detector acceptance.
We note that, as remarked in  \cite{Lipari-Lusignoli},
the  detectors do not  have a sharp, angle  independent  threshold, and 
therefore the corrections  are in principle  model dependent.

\begin{figure} [t]
\centerline{\psfig{figure=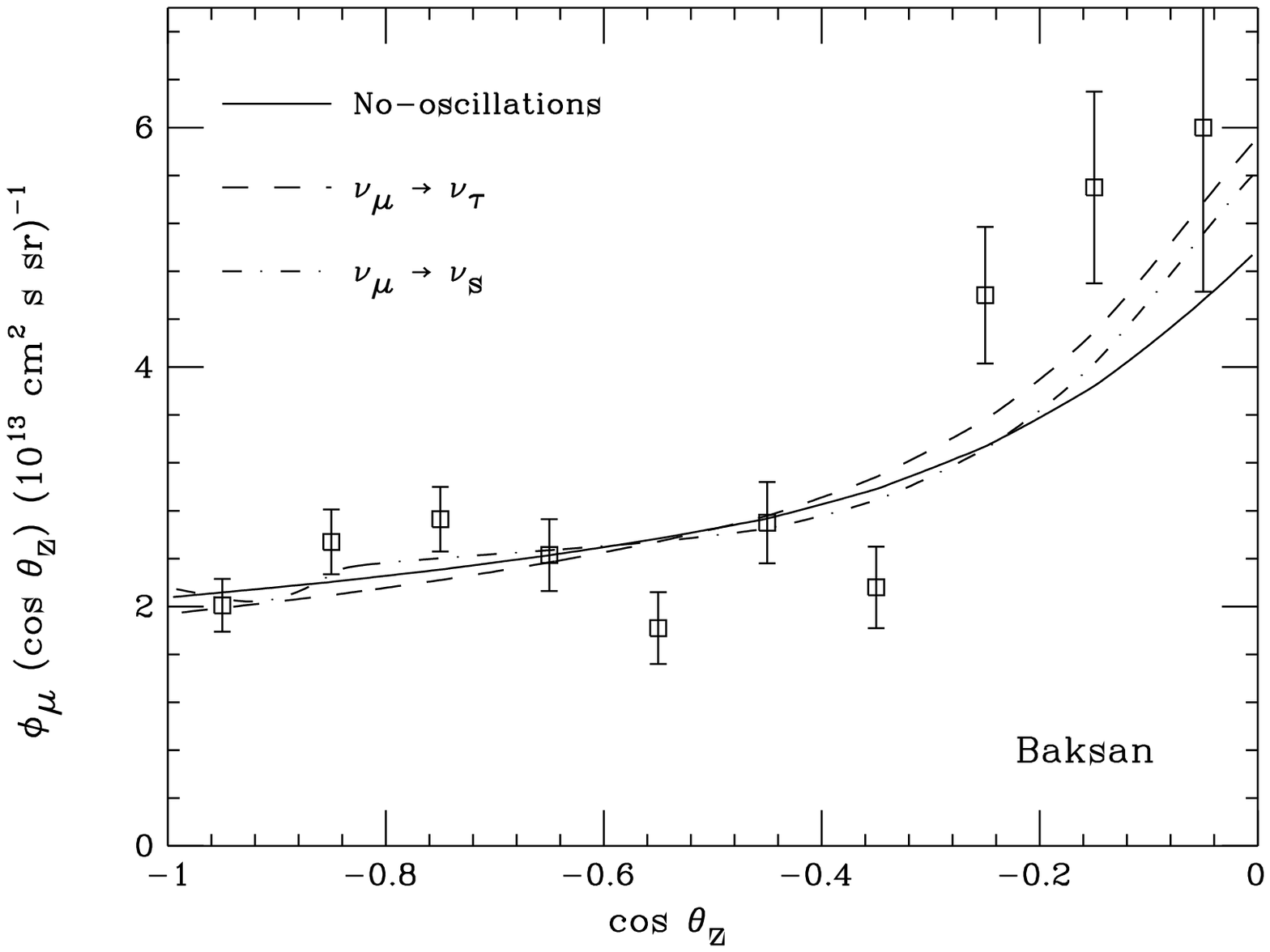,height=9cm}}
\vspace {0.4 cm}
\centerline{\psfig{figure=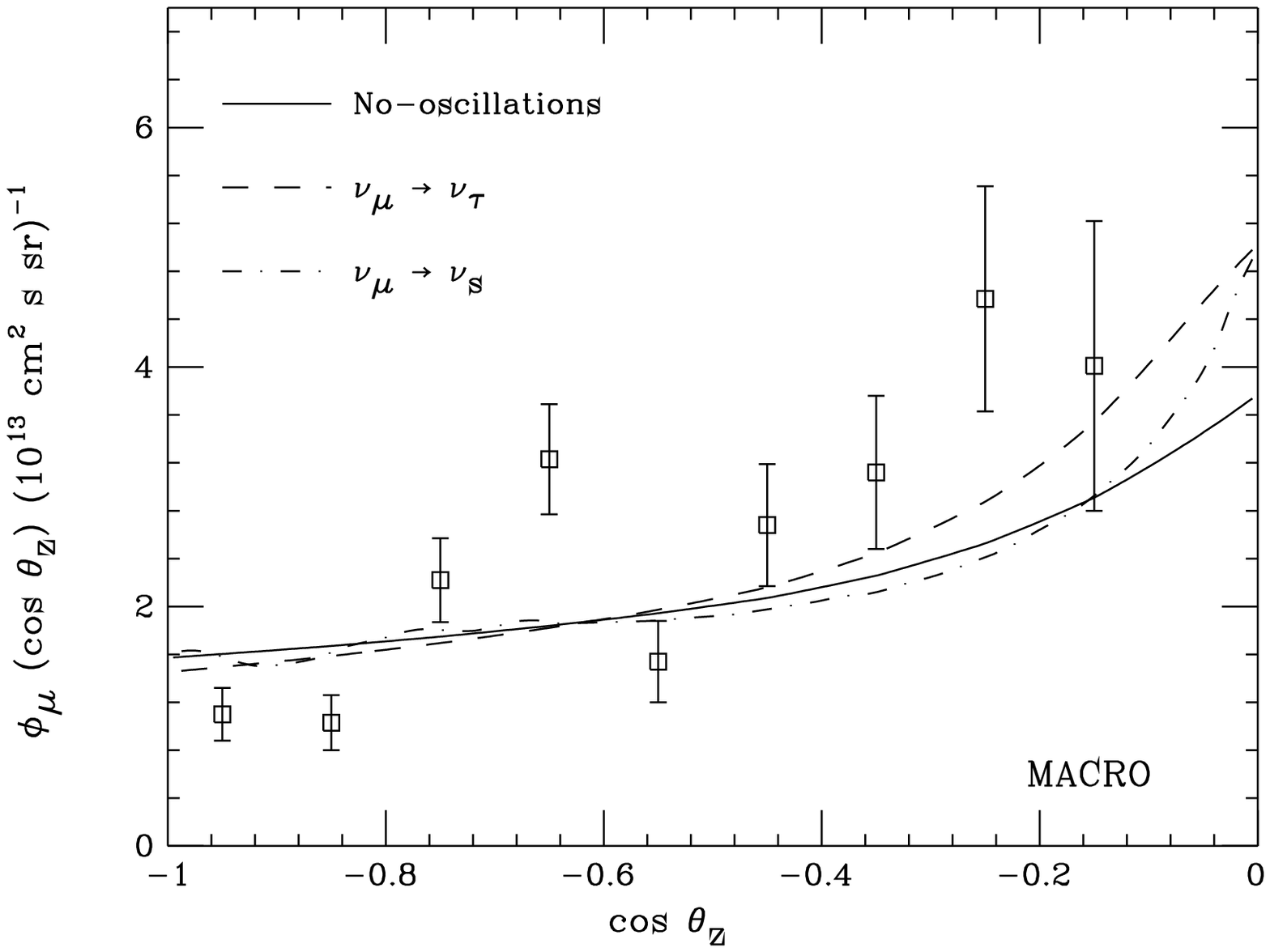,height=9cm}}
\vspace {0.4 cm}
\caption{Flux of  through--going upward going muons  measured  in 
the  Baksan  \protect\cite{Baksan-upmu} (upper  panel)
and MACRO  \protect\cite{MACRO-upmu} (lower  panel) detectors,  
compared  to    the no--oscillation  hypothesis  and to  
oscillation  models for  $\nu_\mu$--$\nu_\tau$  and  $\nu_\mu$--$\nu_s$   
with best fit  parameters (see text).
\label{fig:Baksan-MACRO}
}
\end{figure}

We present in fig.~\ref{fig:Baksan-MACRO} the data of
the Baksan and MACRO detectors, together with fitted
theoretical predictions. The fits are quite poor. 

For Baksan the $\chi^2$  of the no--oscillation
hypothesis  is  $\chi^2 \simeq 26.6$  with a corresponding 
normalization factor  $\alpha \simeq 0.95$.
Including $\nu_\mu \leftrightarrow \nu_\tau$
($\nu_\mu \leftrightarrow \nu_s$)  oscillations does  not  result in a  
significantly better fit to the data:
the minimum $\chi^2$ is 25.7 (23.2)    for 
$\Delta m^2 \simeq 1.0\cdot 10^{-3}$~eV$^2$, $\sin^2 2\,\theta \simeq 0.70$
and $\alpha = 1.17$
($\Delta m^2 \simeq -1.1\cdot 10^{-3}$~eV$^2$, $\sin^2 2\,\theta \simeq 0.90$,
$\alpha = 1.08$).

For MACRO  the $\chi^2$  for the  no oscillation
hypothesis  (considering  only  statistical errors) is
$\chi^2 \simeq 39$    (with $\alpha \simeq 0.70$).
Including $\nu_\mu \leftrightarrow \nu_\tau$  
($\nu_\mu \leftrightarrow \nu_s$)   oscillations
the $\chi^2$ is  minimized  ($\chi^2 = 30.2$  and 35.5)  
with  the  choice  of parameters:
$\Delta m^2 \simeq 1.5\cdot 10^{-3}$~eV$^2$, $\sin^2 2\,\theta \simeq 1$,
$\alpha = 0.97$ 
($\Delta m^2 \simeq 1.5\cdot 10^{-2}$~eV$^2$, $\sin^2 2\,\theta \simeq 1$,
$\alpha = 1.02$). 
In the $\nu_\mu \leftrightarrow \nu_\tau$    hypothesis  the 
$\chi^2$  is reduced   by  7 units, however given its large value
the  significance of the result  is not  clear.

The  zenith angle distribution  of the MACRO  data  shows some
unexplained structure  with a deficit for directions close
to the vertical and  an  excess for 
$\cos\theta_z \simeq -(0.6$--0.7).
Liu  and  Smirnov \cite{Liu-Smirnov} have  suggested that this  structure 
could  be  evidence for oscillations of muon into sterile neutrinos, 
observing that   the  qualitative  features of the  measured  angular
distribution  are  similar to those  of the asymptotic form  of the 
oscillation   probability (see fig.~3).
A more detailed   calculation  does  not  support this claim.  
In the case of  $\nu_\mu$--$\nu_s$  mixing  the  matter  effects  do
predict   features   like  a  reduction  of the flux for
$\cos \theta_z \sim -0.9$,  however the  amplitude of  the  effect  is much
smaller  than  what  is  measured by MACRO.

\section {Conclusions}

The simplest explanation for the atmospheric neutrino anomaly
is the transition of muon neutrinos and antineutrinos into
tau or sterile (anti)neutrinos.  If the muon neutrinos
are mixed with a singlet sterile state,  the  
energy and  zenith angle dependence of the oscillation probability
is  modified by matter effects.  The differences
between  the cases 
of $\nu_\mu$--$\nu_\tau$  and  $\nu_\mu$--$\nu_s$ mixing
are important for  $E_\nu/|\Delta m^2| \aprge 10^3$~GeV/eV$^2$.
The two cases are distinguishable with 
high statistics studies of multi--GeV events, but only  
for the lower range of values of $|\Delta m^2|$ allowed by the analysis of
(semi--)contained events. For larger
$|\Delta m^2|$, the distinction will be possible through a
detailed study of upgoing muons. The muon data  available at  present
are of difficult interpretation, in that only the data obtained by the
Super--Kamiokande collaboration are in agreement with theoretical 
expectations. Moreover, the Baksan and MACRO experimental data disagree with
each other.
The SK data favour neutrino oscillations, albeit with
low statistical significance; both 
 $\nu_\mu \leftrightarrow \nu_s$ (with slightly better agreement) 
 and  $\nu_\mu \leftrightarrow \nu_\tau$
oscillations provide good fits. We can expect that forthcoming 
additional data will clarify the situation and allow to identify the
right solution. 

\vspace {0.5 cm}  
\noindent {\bf Acknowledgments} 

\vspace {0.2 cm}
The authors  wish  to  acknowledge  useful  discussions with S.P.Mikheyev.

\newpage

\begin{thebibliography}{999}
\bibitem {SK-sub}
K.S.\ Hirata {\em et al.}, Phys.Lett. B {\bf 205}, 416 (1988) and 
{\em ibid.} {\bf 280}, 146 (1992); \\
Y.\ Fukuda {\em et al.}, Phys.Lett. B {\bf 335}, 237 (1994); \\
D.\ Casper {\em et al.}, Phys.Rev.Lett. {\bf 66}, 2561 (1991); 
R.\ Becker--Szendy {\em et al.}, Phys.Rev. D {\bf 46}, 3720 (1992);\\
T.\ Kafka, Soudan--2 collaboration, Proceedings of TAUP97; \\
Y.\ Fukuda {\em et al.}, Super--Kamiokande collaboration, 
hep-ex/9803006, submitted to Phys.Lett.

\bibitem {SK}
E.\ Kearns, Proceedings of TAUP97, hep-ex/9803007.

\bibitem{solar} see for example V.\ Berezinsky, 
Invited lecture at 25$^{th}$ ICRC,
Durban, July 28 - August 8, 1997, astro-ph/9710126.

\bibitem{LSND} C. Athanassopoulos et al, (LSND Collaboration)
Phys. Rev. Lett. {\bf 77}, 3082 (1996);  nucl-ex/9706006; nucl-ex/9709006.

\bibitem{three-neutrinos}
There  have  been several attempts  to  reconcile all
detected anomalies  with  a  scheme  involving only 
three neutrino flavors,
suggesting   for  example  that
$\Delta m_{LSND}^2 = \Delta m_{atm}^2$ 
\protect\cite{Cardall-Fuller,minimum-sacrifice},
or  that $\Delta m_{\odot}^2 = \Delta m_{atm}^2$ 
\protect\cite{Acker-Pakvasa}.
The  first  scenario  predicts the 
absence  of a zenith angle dependence of the 
atmospheric neutrino   anomaly, 
the second one  predicts a constant suppression for all solar neutrino
experiments and,
since the atmospheric  neutrino  anomaly is essentially due
to  $\nu_\mu \leftrightarrow \nu_e$ in this case, a positive  signal  for the
Chooz  experiment \protect\cite{Chooz};  therefore both scenarios  are 
currently not  supported by the data.

\bibitem{Cardall-Fuller} C.Y.Cardall and  G M.Fuller,
Phys. Rev. D {\bf 53}, 4421 (1996).

\bibitem{minimum-sacrifice}
G.L. Fogli, E. Lisi, D. Montanino  and G. Scioscia,
Phys. Rev. D {\bf 56}, 4365 (1997).

\bibitem{Acker-Pakvasa} A. Acker and S. Pakvasa, 
 Phys. Lett. B {\bf 397}, 209  (1997).

\bibitem {Chooz} Chooz collaboration: M. Apollonio {\it et al.},  
 hep-ex/9711002.

\bibitem{LEP} The current  value for the number of 
standard light neutrinos  is $N_\nu = 2.995 \pm 0.061$, see
LEP Electroweak Working Group, CERN/PPE/95--172.
 
\bibitem{Concha98} M.C.\ Gonzalez--Garcia, H.\ Nunokawa, O.L.G.\ Peres,
T.\ Stanev and J.W.F.\ Valle, hep-ph/9801368.

\bibitem{Foot-Volkas-numu-nue}
R. Foot, R. R. Volkas and O. Yasuda, hep-ph/9802287.

\bibitem{ALL93} 
E.\ Akhmedov, P.\ Lipari, and M.\ Lusignoli,
Phys.\ Lett.\ B {\bf 300}, 128 (1993).  

\bibitem{ster_other} R.\ Foot, Mod.\ Phys.\ Lett.\ A {\bf 9}, 169 (1994);\\
R.\ Foot and R.R.\ Volkas, Phys.\ Rev.\ D {\bf 52}, 6595 (1995);\\
S.M.\ Bilenky, C.\ Giunti and W.\ Grimus, Proc. of {\it Neutrino 96},
Helsinki, June 1996, edited by K.\ Enqvist {\it et al.}, p. 174.

\bibitem{Vissani-Smirnov} F.\ Vissani and A.Yu.\ Smirnov, hep-ph/9710565.

\bibitem{Liu-Smirnov} Q.Y.\ Liu and  A.Yu.\ Smirnov, hep-ph/9712493.

\bibitem{Foot_etal} R.\ Foot, R.R.\ Volkas and O.\ Yasuda, hep-ph/9801431.


\bibitem{cosmological-bounds}
A. Dolgov, Sov. J. Nucl. Phys. {\bf 33}, 700 (1981);\\
R. Barbieri and A. Dolgov, Phys. Lett. B {\bf 237}, 440 (1990) and
Nucl. Phys. B {\bf 349}, 743 (1991);\\
K. Enqvist, K. Kainulainen and M. Thomson, 
Phys. Rev. Lett. {\bf 68}, 744 (1992) and 
Nucl. Phys. B {\bf 373}, 498 (1992);\\
J. Cline, Phys. Rev. Lett. {\bf 68}, 3137 (1992);\\
X. Shi, D. N. Schramm and B. D. Fields, Phys. Rev. D {\bf 48}, 2568 (1993).
  
\bibitem {Schramm-Turner}
D.Schramm and M.Turner, Rev. Mod. Phys. {\bf 70}, 303 (1998).


\bibitem{Foot-Volkas} R.Foot and  R.Volkas, 
Phys. Rev. D  {\bf 55}, 5147  (1997);\\
R.Foot and R.Volkas, Phys. Rev. D {\bf 56}, 6653 (1997).

\bibitem{NC-exp} S.J. Barish {\it et al.},  Phys.Rev.D {\bf 19}, 2521 (1979).

\bibitem{NC-theo} G.L. Fogli and G. Nardulli, 
Nucl.Phys.B {\bf 160}, 116 (1979).

\bibitem{MSW} L.\ Wolfenstein, Phys. Rev. D {\bf 17}, 2369 (1978);\\
S.P.\ Mikheyev and A.Yu.\ Smirnov, Yad. Fiz. {\bf 42}, 1441 (1985) 
[Sov.J.Nucl.Phys. {\bf 42}, 913 (1985)].

\bibitem{footnote-parametric}  
The  enhancement of  the amplitude  of  oscillations
at $\cos\theta_z \simeq 0.9$,
in the case of $\nu_\mu$--$\nu_s$ mixing, 
can be understood \protect\cite{Liu-Smirnov}
as  a  case of `parametric resonance'~\protect\cite{parametric-res}.

\bibitem{parametric-res} E.\ Akhmedov, Yad.Fiz. {\bf 47}, 475 (1988); \\
P.I.\ Krastev and A.Y.\ Smirnov, Phys. Lett. B {\bf 226}, 341 (1989).

\bibitem {Stacey} F.D. Stacey, ``Physics of the earth'', Wiley, New York, 
1969.

\bibitem {footnote-density} The  dashed  curves 
in fig.~\protect\ref{fig_prob_a} and \protect\ref{fig_prob_b} are
calculated  assuming  a  constant  density in the core  and 
mantle of the earth $\langle \rho_c (\theta_z)\rangle$
and $\langle \rho_m (\theta_z)\rangle$, calculated  averaging the density
along a trajectory of zenith angle $\theta_z$.

\bibitem {footnote-angular} Note in the lower  panels of
fig.~\protect\ref{fig_ang1} and~\protect\ref{fig_ang2}  that the
suppression factor  for  $\nu_\mu \to \nu_s$ transitions,
if studied in great  detail shows  a   complex  dependence
on the zenith angle  that  can be 
compared with the oscillation  probability curves 
in fig.~\protect\ref{fig_prob_a}.

\bibitem {nu-flux-comp} T. K. Gaisser, M. Honda, 
K. Kasahara, H. Lee, S. Midorikawa, V. Naumov, T. Stanev,
Phys.Rev. D {\bf 54},  5578  (1996).

\bibitem{kamioka-semi}
Y.\ Fukuda {\em et al.}, Phys.Lett. B {\bf 335}, 237 (1994).

\bibitem{footnote-Foot}
We note that in 
\protect\cite{Foot_etal}  the authors  have  considered   only the case 
of positive $\Delta m^2$.  
Matter  effects allow  to  distinguish the case  of positive  and  negative
squared  mass differences, and  the allowed  region  for $\Delta m^2 < 0$ 
is  somewhat larger than in the other  case, because of the  MSW  resonance
present  for the neutrinos.

\bibitem{Bartol}  V.\ Agrawal, T.\ K.\ Gaisser, P.\ Lipari, and T.\ Stanev,
Phys.\ Rev.\ D {\bf 53} 1314, (1996).

\bibitem {LLS}  P.\ Lipari, M.\ Lusignoli, and F.\ Sartogo,
		Phys.Rev.Lett. {\bf 74}, 4384 (1995).

\bibitem {nu-position}  T.K. Gaisser and Todor Stanev, 
Phys. Rev. D {\bf 57}, 1977 (1998).


\bibitem {nu-induced-muons}
The flux of $\nu$--induced muons  is  larger   than the
flux of  atmospheric  muons  for line  of  sights 
shielded by a column  density larger  than $\sim  10^6$~g~cm$^{-2}$.
In the upper hemisphere this condition is satisfied  
at most in  a small solid angle for the deepest  detectors.

\bibitem{Lipari-Lusignoli}
P.\ Lipari and M.\ Lusignoli, hep-ph/9712278,
Phys. Rev. D {\bf 57}, 3842R (1998).

\bibitem {Lohmann} W. Lohmann, R. Kopp \& R. Voss, CERN Yellow Report
No. EP/85-03.

\bibitem{Baksan-upmu}
Baksan Collaboration:
M.\ M.\ Boliev {\it et al.},
ICRC '95, Rome,
Vol~1, p.~686; {\em ibidem}, p.~722.

\bibitem{MACRO-upmu}
MACRO Collaboration: S.\ Ahlen {\em et al.},
Phys.\ Lett.\ B {\bf 357}, 481 (1995); 
F.\ Ronga {\em et al.}, in  
{\em Neutrino '96}, Proceedings of the 17th Conference
on Neutrino Physics and Astrophysics, Helsinki, 1996.

\bibitem{Kam-upmu}
Kamiokande Collaboration: A.\ Suzuki {\em et al.}
in Proceedings of 7th International Workshop on Neutrino 
Telescopes, Venice, 1996, edited by M.\ Baldo Ceolin
 p.~263.
 
\bibitem{IMB-upmu}
IMB Collaboration: R.\ Becker-Szendy {\em et al.},
Phys.\ Rev.\ Lett.\ {\bf 69}, 1010 (1992); \\
D.\ W.\ Casper, in 
Proceedings of 3rd International Workshop on Neutrino 
Telescopes, Venice, 1991, edited by M.\ Baldo Ceolin  p.~213.

\bibitem{HKKM}
M.\ Honda, T.\ Kajita, K.\ Kasahara, and S.\ Midorikawa,
Phys.\ Rev.\ D {\bf 52}, 4985 (1995);

\bibitem{GRV}
M. Gluck, E. Reya, and A. Vogt,
Z.Phys. C {\bf 67}, 433 (1995).

\bibitem{Frati93}
W.\ Frati, T.\ K.\ Gaisser, A.\ K.\ Mann, and T.\ Stanev,
Phys.\ Rev. D {\bf 48}, 1140 (1993).
 

\bibitem{note_area} The acceptance for the muon flux increases with the area
(as opposed to the volume) of the detector: SK does not dominate the other
experiments as for the contained neutrino events.
 
\bibitem{Fogli-upmu}
G.L. Fogli, E. Lisi and A. Marrone,
preprint BARI-TH-280-97, hep-ph/9708213.

\end{thebibliography}
\end{document}